\documentclass[proof]{pasj00}
\draft

\begin{document}
\SetRunningHead{Author(s) in page-head}{Running Head}
\Received{ / /}%{yyyy/mm/dd}
\Accepted{ / /}%{yyyy/mm/dd}

\title{Chemical composition of carbon-rich, very metal-poor subgiant LP625-44 observed with the Subaru/HDS\thanks{Based on data collected at Subaru Telescope, which is operated by the National Astronomical Observatory of Japan. }}

%%% begin:list of authors
%\author{Wako \textsc{Aoki}}%
%  \thanks{Example: Present Address is xxxxxxxxxx}}
%\affil{National Astronomical Observatory, Mitaka, Tokyo, 181-8588 Japan}
%\email{aoki.wako@nao.ac.jp}
%\author{Hiroyasu \textsc{Ando}}%
%%  \thanks{Example: Present Address is xxxxxxxxxx}}
%\affil{National Astronomical Observatory, Mitaka, Tokyo, 181-8588 Japan}
%\email{ando@subaru.naoj.org}

%\author{B-Firstname \textsc{B-Familyname}}
%\affil{B-Address of Institute}\email{bbbbb@xxx.xxx.xx.xx}
%\and
%\author{C-Firstname {\sc C-Familyname}}
%\affil{C-Address of Institute}\email{ccccc@xxx.xxx.xx.xx}
%%% end:list of authors

%%% Please use the following style in case that sorting by 
%%% affilation is impossible. 
%
 \author{%
   Wako \textsc{Aoki}\altaffilmark{1},
   Hiroyasu \textsc{Ando}\altaffilmark{2},
   Satoshi \textsc{Honda}\altaffilmark{1},
   Masanori \textsc{Iye}\altaffilmark{1},
   Hideyuki \textsc{Izumiura}\altaffilmark{3}, \\ 
   Toshitaka \textsc{Kajino}\altaffilmark{1}, 
   Eiji \textsc{Kambe}\altaffilmark{4},
   Satoshi \textsc{Kawanomoto}\altaffilmark{1},
   Kunio \textsc{Noguchi}\altaffilmark{1}, \\
   Kiichi \textsc{Okita}\altaffilmark{1}, 
   Kozo  \textsc{Sadakane}\altaffilmark{5}, 
   Bun'ei \textsc{Sato}\altaffilmark{3,6},
   Ian \textsc{Shelton}\altaffilmark{2},
   Masahide \textsc{Takada-Hidai}\altaffilmark{7}, \\
   Yoichi \textsc{Takeda}\altaffilmark{8}, 
   Etsuji \textsc{Watanabe}\altaffilmark{3}
     and  
   Michitoshi \textsc{Yoshida}\altaffilmark{3}} 
%   and
%   F-Firstname \textsc{F-Familyname}\altaffilmark{2}}

 \altaffiltext{1}{National Astronomical Observatory, Mitaka, Tokyo, 181-8588}
 \email{aoki.wako@nao.ac.jp}
% \email{honda@optik.mtk.nao.ac.jp}
% \email{kajino@th.nao.ac.jp}
% \email{kawanomo@optik.mtk.nao.ac.jp}
% \email{knoguchi@optik.mtk.nao.ac.jp}
% \email{okita@optik.mtk.nao.ac.jp}
 \altaffiltext{2}{Subaru Telescope, National Astronomical Observatory of Japan, \\
650 North Aohoku Place, Hilo, HI96720, USA}
% \email{ando@subaru.naoj.org}
 \altaffiltext{3}{Okayama Astrophysical Observatory, National Astronomical Observatory of Japan, \\
Kamogata-cho, Okayama, 719-0232}
% \email{izumiura@oao.nao.ac.jp}
% \email{satobn@oao.nao.ac.jp}
% \email{watanabe@oao.nao.ac.jp}
 \altaffiltext{4}{Department of Earth and Ocean Sciences, National Defense Academy, \\
Hashirimizu 1-10-20, Yokosuka, Kanagawa 239-8686}
% \email{kambe@nda.ac.jp}
 \altaffiltext{5}{Astronomical Institute, Osaka Kyoiku University, \\
Kashiwara-shi, Osaka, 582-8582}
% \email{sadakane@cc.osaka-kyoiku.ac.jp}
 \altaffiltext{6}{Department of Astronomy, School of Science, University of Tokyo, \\
Bunkyo-ku,Tokyo 113-0033}
 \altaffiltext{7}{Liberal Arts Education Center, Tokai University, 1117 Kitakaname, \\
Hiratsuka-shi, Kanagawa, 259-1292}
% \email{hidai@apus.rh.u-tokai.ac.jp}
 \altaffiltext{8}{Komazawa University, Komazawa, Setagaya, Tokyo, 154-8525}
% \email{takedayi@cc.nao.ac.jp}
% \email{eeeee@xxx.xxx.xx.xx}
% \altaffiltext{2}{Address of Institute}

%% `\KeyWords{}' always has to be placed before `\maketitle'.
\KeyWords{nuclear reactions, nucleosynthesis, abundances -- stars: abundances -- stars: AGB and post-AGB -- stars: carbon -- stars: Population II} %Do NOT move this preamble from here!

\maketitle

\begin{abstract}

We have obtained a high resolution ($R\sim 90,000$) spectrum of the
carbon- and {\it s}-process-element-rich, very metal-poor
([Fe/H]$=-2.7$) subgiant LP~625-44, as well as that of HD~140283 (a
metal-poor subgiant with normal abundance ratio) for comparison, with
the High Dispersion Spectrograph (HDS) for the Subaru Telescope for
detailed abundance study. The spectrum covers 3400-7800{\AA} and
enables us to study important spectral lines which had not been
detected in the previous works. We found significant overabundances in
some light elements, in addition to carbon and nitrogen for which
large enhancements were already known.  The oxygen abundance derived
from the O {\small I} triplet around 7770{\AA} is uncertain, but the
excess of oxygen in LP~625-44 seems remarkable (perhaps by nearly a
factor 10), in comparison with that of HD~140283 derived from the same
lines. The Na enhancement in LP~625-44 is by about a factor 50,
suggesting hydrogen burning in the $^{22}$Ne-rich layer in an
asymptotic giant branch star which produces the abundance pattern of
this object. In our new spectrum of LP~625-44, the Pb {\small I}
$\lambda$3683{\AA} line has been detected, as well as the Pb {\small
I} $\lambda$4057{\AA} line which has already been studied, confirming
the Pb abundance ($\log \epsilon$(Pb)$\sim 1.9$) derived by the
previous work. The abundance ratio of {\it s}-process elements at the
second peak (e.g., La, Ce and Nd) to that at the third peak (Pb) in
LP~625-44 is significantly higher (by a factor 5) than that in other
three {\it s}-process element-rich objects recently studied by van Eck
et al.. Recent theoretical works have modeled the {\it s}-process
nucleosynthesis in the radiative layer of asymptotic giant branch
stars in the inter-pulse phase, and the above results means that these
processes produced a large scatter in the abundance ratios. Another
possiblity is that different processes (e.g., {\it s}-process
nucleosynthesis during thermal pulses) have contributed to heavy
elements in the early Galaxy.

\end{abstract}

\section{Introduction}\label{sec:intro}

The chemical composition of very metal-poor stars (e.g., [Fe/H]$<-2.5$) is
believed to be determined by a small number of nucleosynthesis
processes preceding the formation of these objects. Abundance studies
for these objects in the past decade have provided rich information
for our understanding of individual nucleosynthesis processes in the
early Galaxy \citep{mcwilliam95,ryan96}.

The very metal-poor ([Fe/H]$\sim -2.7$) subgiant LP~625-44 shows
extremely large overabundances of carbon, nitrogen and neutron-capture
elements \citep{norris97a}. This abundance property indicates the
nucleosynthesis products in the asymptotic giant branch (AGB) stage of
low-mass or intermediate-mass stars ($M=1 \sim 8 M_{\odot}$, Iben \&
Renzini 1983). Since the evolutionary time-scale of these objects is
long ($100$M$ \sim$1G year), the abundance pattern of LP~625-44 is not
considered to originate in the gas cloud from which this object
formed, but to reflect the yields of nucleosynthesis after this object
has formed. One possible model to explain the abundances of this
object is that involving mass transfer from a carbon-enhanced AGB star
(one that has since evolved to the white dwarf stage and cannot now be
seen) to its lower-mass companion, which is now observed as a
carbon-rich object like LP625-44. The binarity of this object was
confirmed by radial-velocity monitoring \citep{aoki00}, strongly
supporting the above mass transfer scenario. This means that the
nucleosynthesis process even in low-mass and intermediate-mass stars
can also be studied by abundance analyses of such special metal-poor
objects. The abundance pattern of the heavy neutron capture elements
in LP~625-44 agrees well with that of solar system {\it s}-process
elements \citep{aoki00}. Detailed abundance study of this star will
give a strong constraint on {\it s}-process nucleosynthesis at low
metallicity and AGB evolution in the early Galaxy.

Recent modeling of AGB stars by \citet{straniero95} showed that
neutron capture mainly occurs, not in the convective He shell {\it
during} a thermal pulse, but in the radiative state {\it between} two
given pulses. In this model the density distribution of the
$^{13}$C-rich layer (referred to as the $^{13}$C pocket), which
provides neutrons for the {\it s}-process, is taken as a free
parameter. The yields of neutron-capture elements have been
systematically calculated based on the stellar models by
\citet{gallino98}, who succeeded in reproducing at least the main
component of the solar-system {\it s}-process elements. The {\it
s}-process nucleosynthesis in the inter-pulse phases was also studied
by Goriely \& Mowlavi (2000), who model the mixing process via a
diffusion mechanism. On the other hand, quite recently,
\citet{iwamoto02} have found theoretically that the proton mixing into
helium burning layer occurs {\it during} thermal pulses {\it only in}
very metal-deficient ([Fe/H] $\le -$2.5) AGB stars (see section 4.3
for details). This model predicts a large neutron exposure, and a
different {\it s}-process abundance pattern from those of the standard
model of \citet{gallino98} and \citet{goriely00}.

One issue to be studied for LP~625-44 is the lead (Pb)
abundance. The detection of Pb in LP~625-44 by \citet{aoki00} enables
a comparison of the abundance pattern from Sr ($Z=38$) to Pb ($Z=82$)
with predictions of AGB models. The Pb {\small I} line ($\lambda
4057$) used in the above study has recently been detected in other
three objects by \citet{vaneck01}. An important result of the work is
that the Pb of these three objects are much more abundant than the
elements with $Z\sim 60$ (e.g., Ce and Nd), and the abundance ratios
like Ce/Pb and Nd/Pb are significantly lower than those of
LP~625-44. This implies a large scatter in these abundance ratios
which represent the production ratio of the elements at the second- to
the third-peaks for {\it s}-process at low metallicity. Despite the
importance of its Pb abundance, the abundance analysis for Pb in
LP~625-44 relies only on the $\lambda 4057${\AA} line (the situation
is the same for the other three objects). Confirmation of the Pb
abundance by other lines is strongly desired.

Another issue concerning the chemical composition of LP~625-44 is the
abundances of light elements. {\it S}-process nucleosynthesis is
dependent on the conditions of the burning layer in AGB stars, which
may be investigated by the resulting abundances of carbon, nitrogen,
oxygen and other light elements. A large excess of carbon and
nitrogen, and moderate enhancement of magnesium have been revealed by
\citet{norris97a} and \citet{aoki00}. Other elements like oxygen (O)
and sodium (Na) are also useful to assess the nucleosynthesis
processes which have contributed to the chemical nature of LP~625-44.

For the study mentioned above, we obtained high resolution spectrum of
LP~625-44 with the High Dispersion Spectrograph (HDS) of the 8.2m
Subaru Telescope. The HDS is an \'{e}chelle spectrograph for the
optical wavelength region (3000-10,000{\AA}) located at one of the
Nasmyth foci of the telescope. HDS successfully achieved its
first-light in early July, 2000. Our spectra have been acquired in the
first and second test observing runs for HDS, and show the quality of
the spectrograph, though some minor problems still existed in the
early phases of performance verification.

In this paper, we study the characteristics of the abundances of the
carbon- and {\it s}-process-element-rich star LP~625-44, comparing
with the abundances of the metal-poor subgiant (with normal abundance
ratios) HD140283, which was also observed with similar setups of
HDS. We describe the observation, data reduction and calibration in
some detail in section \ref{sec:obsana}. The equivalent widths,
intrinsic line widths and radial velocities are measured in that
section. In section \ref{sec:ana}, the analysis for absorption lines
of the {\it s}-process elements including Pb {\small I} $\lambda$3683
and those of some light elements are described. We discuss the
interpretation of the large excesses of some light elements including
O and Na, and the variation of abundance patterns for the {\it
s}-process elements in section \ref{sec:disc}.

\section{Observation and Measurements}\label{sec:obsana}

\subsection{Observation with Subaru HDS}\label{sec:obs}

The observation has been made with Subaru/HDS in July and August,
2000. The \'{e}chelle grating of this instrument is a mosaic of two
31.6 gr/mm gratings, and the dispersion is 1~{\AA}/mm at
4300~{\AA}. The detector is a mosaic of two 4k$\times$2k EEV CCD's
with 13.5$\mu$m pixels. We adopted a 0.4~arcsec (0.2~mm) slit width
which enables us to get spectra with resolution about 90,000 by about
3.5 pixels sampling. The high resolving power and high sampling rate
are valuable in the study of the elemental lines affected by the
isotope shifts or by blending with other atomic or molecular lines.

We observed our two objects with both standard blue and red setups of
the spectrograph as given in Table~\ref{tab:obs}. Another spectrum was
also obtained with the UV setup (3100-4500~{\AA}) for HD140283. It
should be noted that the central wavelength range (one order or two)
of each region cannot be observed due to the gap ($\sim$1.1~mm)
between the two CCD's.  While the free spectral range is completely
covered by the blue setup, that is not the case in the wavelength
region longer than 7200~{\AA} in the red setup. A
short-wavelength-cutoff filter (SC46) was used in the observations of
red spectra.

In the observing runs in July and August 2000, the atmospheric
dispersion compensator (ADC) had not yet been installed. The image
rotators were used because the target acquisition and guiding without
image rotators had not yet been tested. In particular, the blue
spectrum (3400-5100~{\AA}) of HD140283 was obtained using the image
rotator for the red setup, because the blue one was not available in
the July run.  We adopted the August data as the blue spectrum
($<$4500~{\AA}) for HD140283. Using the image rotator, the slit image
on the sky was fixed to the pole direction, instead of the zenith
direction, due to the requirement of guiding at that time. Therefore
the loss of light was significant in the short wavelength region in
which the atmospheric differential dispersion is quite large. This
limit in the observation resulted in the lower quality of the spectra
for the short wavelength region ($\lambda < 3300${\AA}).

We note that the ADC for the Nasmyth focus of the Subaru telescope was
installed in January 2001. The throughput is better than that of
image rotators for $\lambda > 3500${\AA}, and observing efficiency has
improved compared to that in the observing runs when we obtained the
data studied in this work. Guiding without image rotators also
became available after our observing runs.

For flat fielding of the CCD data, we obtained Halogen lamp data
(hereafter flat data) with the same setup as that for object
frames. However, we found in the July data that there was a
significant disagreement of the continuum profile and the fringe
pattern between the object spectrum and the flat one. The disagreement
of the continuum profile is not serious for our abundance study
because the analysis is usually made for normalized spectra, but the
difference of the fringe pattern, which is remarkable in the red and
near infrared ranges, directly affects our study of weak absorption
features. By an investigation after the July run, we found that this
problem was mainly due to the mis-alignment of the calibration
system. In the August run, we fixed the problem and obtained the flat
data again with the same setup as in the July run. Using the new flat
data, the calibration quality has significantly improved, and the
fringes have almost disappeared in our spectra.\footnote{There is
still a problem of fringes and continuum profile in HDS data, but that
is not due to the problem of the calibration system, but may be due to
the fluctuation in the angle of the incident beam from the telescope.}
We note that the stability of the detector system including
temperature control is fairly good, and the flat data obtained in
different observing runs are applicable to data with the quality
achieved in our study.

\subsection{Data reduction}\label{obs:red}

In the present data acquisition system of HDS, two FITS formatted CCD
data files are produced for one exposure corresponding to the two
CCD's. These two files are reduced separately in this work. Standard
data reduction using the IRAF\footnote{IRAF is distributed by the
National Optical Astronomy Observatories, which is operated by the
Association of Universities for Research in Astronomy, Inc. under
cooperative agreement with the National Science Foundation.} {\it
echelle} package was applied to each frame following the procedures
mentioned below.

HDS data frames include a so-called ``over-scan region'' which
represents the bias level of the frame. Though the time variation of
the bias level is not large in HDS (1-2 e$^{-}$), we corrected the
bias level by subtracting the average of the counts in the over-scan
region for each frame. Since there are two output points for each CCD,
and the gains (conversion factors) are slightly different between
them, the conversion from ADU to the number of electrons was made
corresponding to the output point.

The dark current of the CCD is about 1e$^{-}$ per hour, and is
negligible in this study. However, some emission inside the Nasmyth
enclosure or leakage of light from the outside about 10~e$^{-}$ per
hour is known. This makes an almost homogeneous background in our
object frames, and can be corrected by the background subtraction
mentioned below. For this reason we have not acquired so-called dark
frames in our observing runs.

Most cosmic-ray noise is excluded by comparing frames obtained by the
same setup of the instrument. When more than three frames were
acquired with the same setup, we replaced the count of a pixel by the
median value of the all frames if the count of the individual frames
is significantly higher than that of the median frame. When only two
frames are available for the identical setup, we made a difference of
the two frames, and replace the count of a pixel by that in another
frame if the difference is significantly larger or smaller than zero.
 
There are several distinctive so-called bad (or hot) columns in the
CCD's, which align with the \'{e}chelle dispersion direction. We
excluded the wavelength regions of the spectra affected by these
columns in the present analysis.

The background subtraction was done by the '{\it apscatter}' routine
of IRAF. The background level is about 2\% of the stellar flux at
4000{\AA} for LP625-44, and higher at shorter wavelength due to the
decrease of the stellar flux. This background level is explained by
reflections of the stellar light at the surface of the CCD and the
field flattener lens just before the CCD. A comparison of the count
level on the slit image and that of the inter-order region has shown
that the sky contamination is negligible for our objects.

%A significant problem was
%found in the halogen lamp data for flat-fielding calibration obtained
%in July 2000: the fringe pattern in near infrared region
%($>$7000{\AA}) could not be canceled by the flat-fielding
%calibration. This problem was almost solved by correcting the incident
%angle of the beam of calibration sources before the August run. We
%applied the calibration data obtained in August to flat-fielding of
%July data. 

As mentioned in the previous subsection, the calibration system was
re-aligned after the July run, and the quality of flat-fielding has
improved. However, the fringes in the longest wavelength range
($\sim$7800{\AA}) cannot necessarily be canceled out by flat fielding,
because of the small variation of the fringe pattern amongst the
object data. Fringes can be the most severe source of noise in the
spectrum for the long wavelength range.

The wavelength calibration was done using the Th-Ar spectra obtained
before and/or after the observations of the targets. The measured rms
wavelength error of the weak Th lines is less than 0.01{\AA}. By the
measurement of the FWHM of weak Th lines, we confirmed the spectral
resolution as about 90,000.

The CCD covers the spectrum of each order much wider than the free
spectral range in the UV-blue region. We trimmed the wavelength range
where the count level is too low (typically lower than 1/4 of the peak
of the \'{e}chelle blaze profile) from the spectrum of each order to
reduce the noise included in the region with low-count level.

For continuum specification, we first made blaze profiles of
individual \'{e}chelle orders using the flat data. The stellar spectra
were divided by these blaze profiles, and the continuum was traced
using IRAF '{\it continuum}' routine. The continuum specification is
still difficult for the regions around broad absorption lines (e.g.,
Balmer lines) and strong molecular bands (e.g., CH G-band), but they
are not important for the present analysis focusing on rather weak
lines.

The S/N ratio of each spectrum, simply estimated from the count level,
is given in the last column of Table~\ref{tab:obs}. This value is
the S/N ratio per pixel (0.012{\AA} at 4000{\AA} and 0.018{\AA} at
6000{\AA}) at the peak of the \'{e}chelle blaze profile. The S/N ratio
at the edge of the \'{e}chelle blaze function is lower than the
center. However, the free spectral range is well covered for most
regions, and the S/N ratio is not severely dependent on the line
position in the \'{e}chelle order after combining the spectra for
neighboring orders.

\subsection{Measurement of equivalent widths}\label{sec:mes}

For the standard analysis which was applied to most elements,
equivalent widths were measured by fitting Gaussian profiles to the
absorption lines. We excluded lines which may be significantly
affected by other absorption lines. The blending with other atomic or
molecular lines was checked using the atomic line list by
\citet{kurucz95} and molecular line lists of CH, CN and C$_{2}$
\citep{aoki01}. The spectral lines which are too broad compared to the
lines with similar strength are also excluded to avoid possible
blending with unknown line absorption. The line list including the
equivalent widths measured is given in Table~\ref{tab:ew}.

\begin{figure}
  \begin{center}
    \FigureFile(70mm,70mm){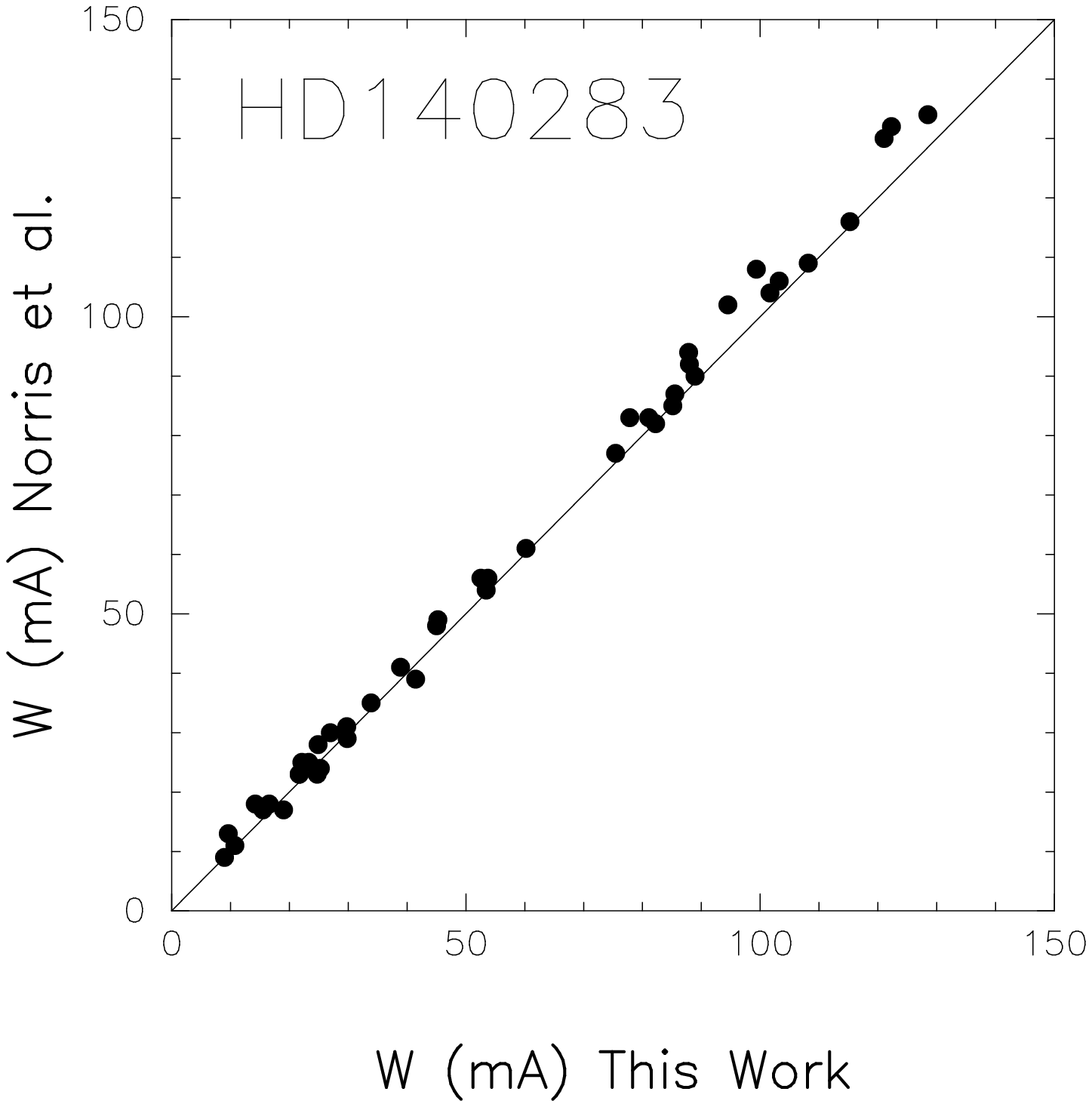}
    \FigureFile(70mm,70mm){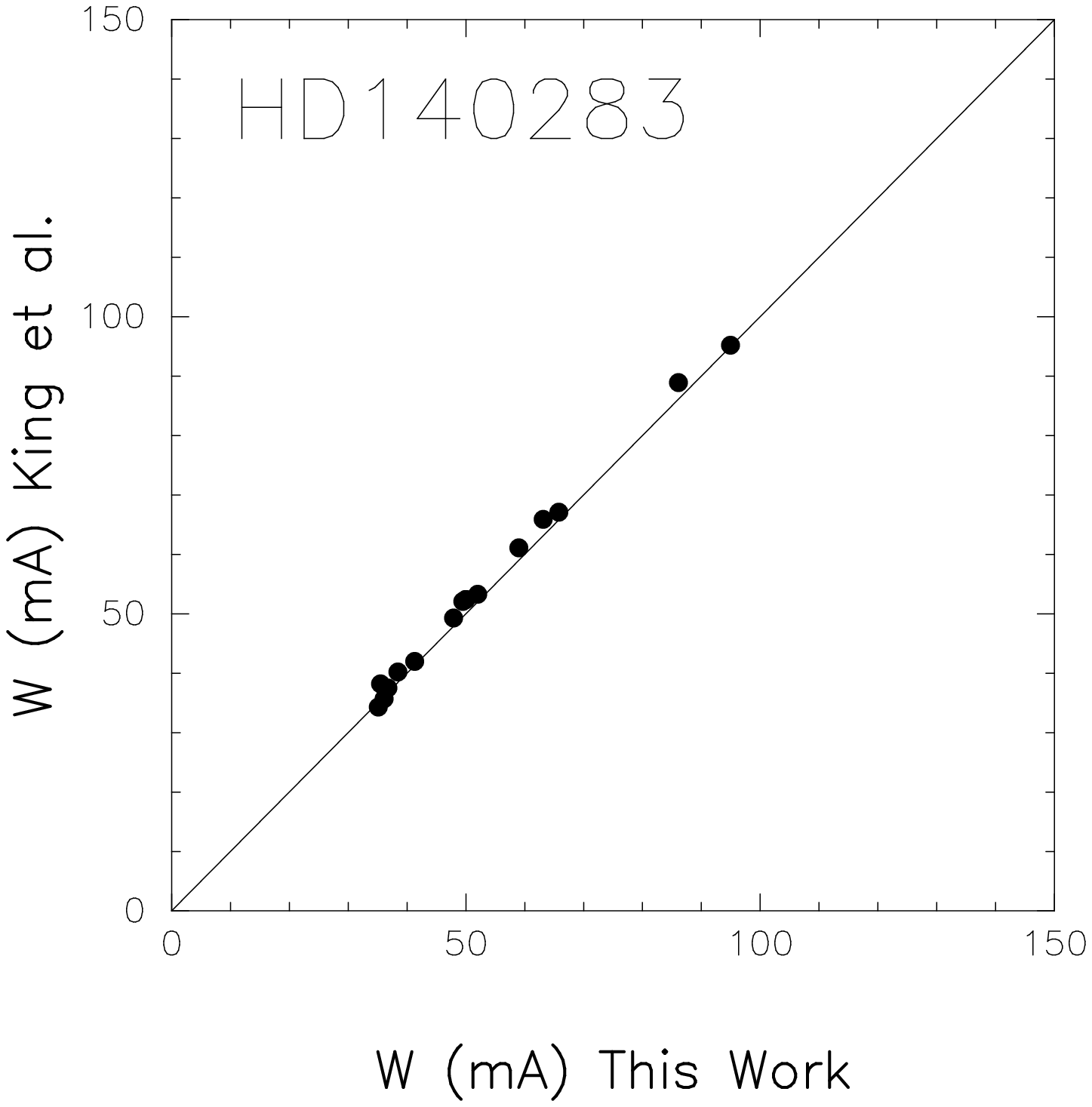}
    \FigureFile(70mm,70mm){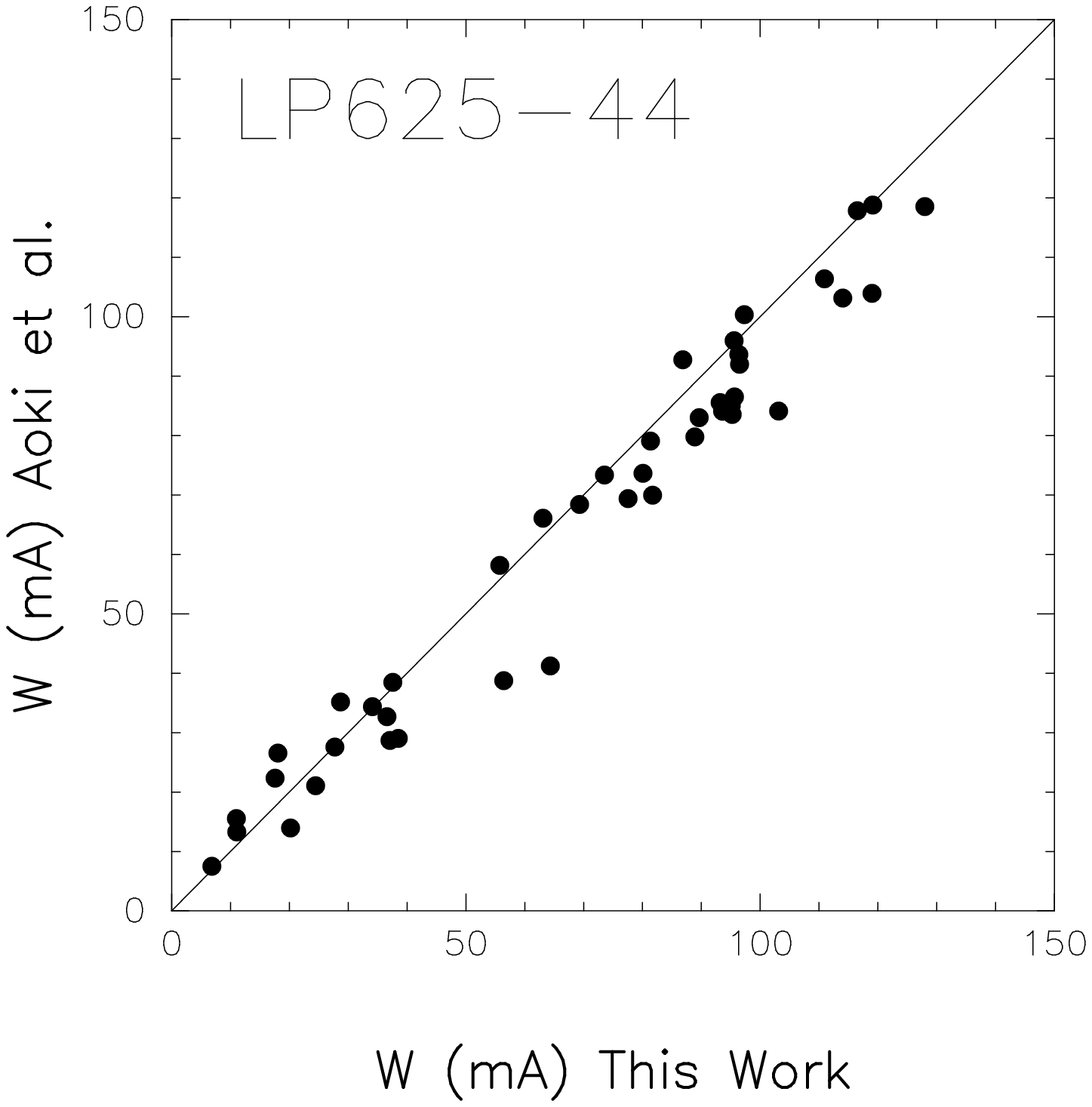}
%%% \FigureFile(width,height){filename}
  \end{center} 

\caption{Comparison of the equivalent widths for Fe {\small I} lines
measured by the present analysis with those of \citet{norris96} and
\citet{king98} for HD140283 and those of \citet{aoki00} for
LP625-44.}

\label{fig:ew}
\end{figure}

In Figure \ref{fig:ew}, the equivalent widths of Fe {\small I}
measured in the present work are compared with those of
\citet{norris96} and \citet{king98} for HD140283, and those of
\citet{aoki00} for LP625-44. The agreement is fairly good between our
measurement and that of \citet{king98}. The agreement between our data
and those by \citet{norris96} is also good for weak lines which are
important for abundance determination, though our equivalent widths of
strong lines are slightly ($\sim$5\%) smaller than theirs. On the
other hand, the dispersion in the diagram for LP~625-44 is larger than
the dispersion in those for HD~140283. This should be due to the lower
S/N ratios of the LP625-44 spectrum than that of the HD140283 one in
both studies, and may also be due to the effect of unidentified
blending, because the lines are very crowded in LP~625-44 due to the
overabundances of carbon, nitrogen and neutron-capture elements.

\subsection{Intrinsic line widths}\label{sec:line}

\begin{figure}
  \begin{center}
    \FigureFile(120mm,140mm){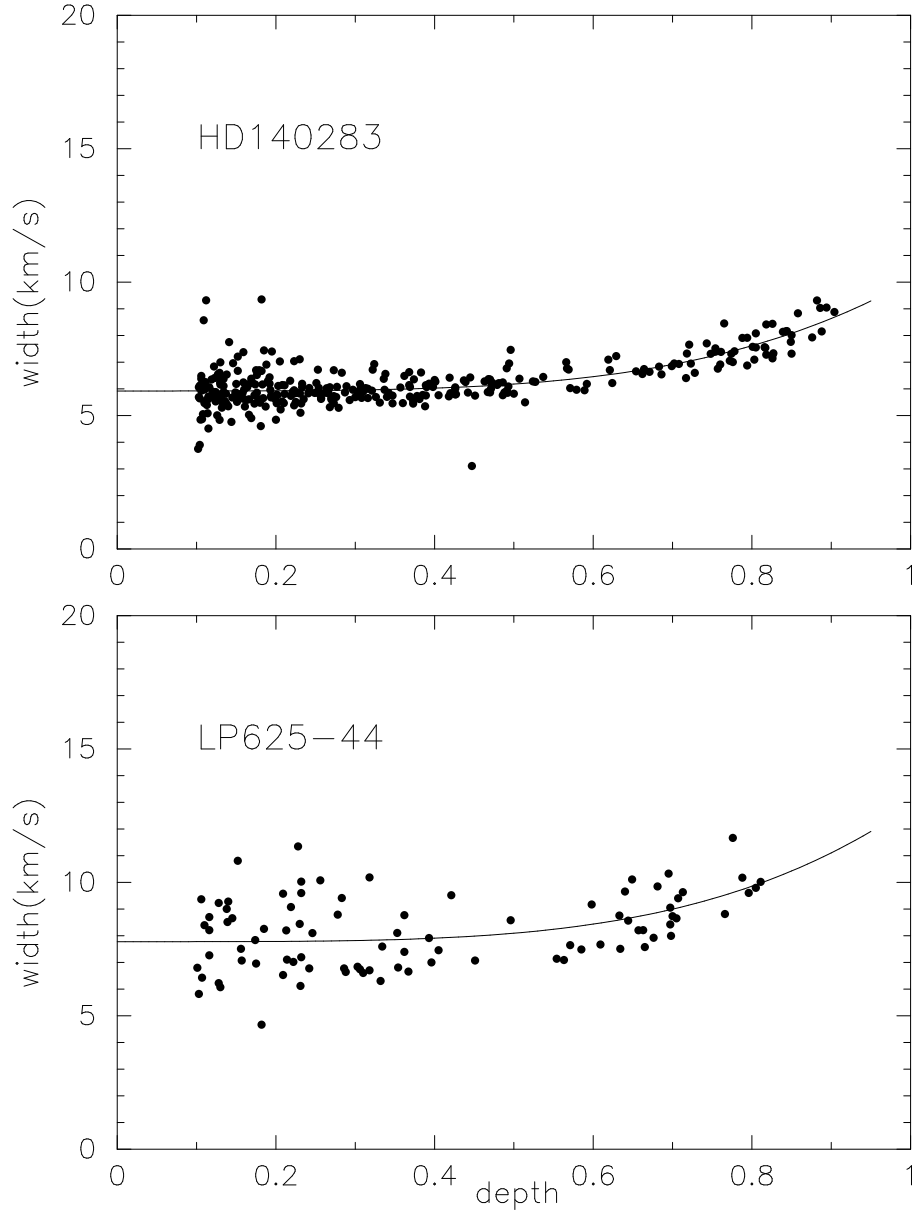}
%%% \FigureFile(width,height){filename}
  \end{center}
 \caption{Line width as a function of depth for Fe 
  {\small I} lines for HD~140283 (upper) and LP~625-44 (lower). Here
  the line width means the Gaussian dispersion in velocity unit (kms$^{-1}$)
  corrected for instrumental effect for which we assumed the
  3.5kms$^{-1}$ (FWHM) Gaussian profile. 
}\label{fig:line}
\end{figure}

Thanks to the high resolving power of HDS ($\sim 3.5~$km~s$^{\small
-1}$), the instrumental line broadening is sometimes smaller than the
intrinsic widths of metal lines due to the motion of the photospheric
gases (macro-turbulence) and/or stellar rotation. We plot the widths
of Fe {\small I} lines observed against their central depths in Figure
\ref{fig:line}. The width in the figure indicates the Gaussian
dispersion in velocity units (km~s$^{-1}$) converted from the FWHM and
corrected for instrumental broadening for which we assumed a Gaussian
profile. We excluded the lines whose central depths are shallower than
0.1 because their quality is not good. Then we determined the
optically thin limit of the line width by extrapolating the plots to
zero central depth. This thin limit can be interpreted as the
intrinsic line width.

The thermal broadening for our objects ($T_{\rm eff}\sim$5500K) is
about 1.8~km~s$^{\small -1}$, and the micro-turbulent velocity is
1.2~km~s$^{\small -1}$ (see section \ref{sec:ana}). Correcting for
these broadening, the macroscopic broadening is estimated to be 5.5
and 7.4~km~s$^{\small -1}$ for HD140283 and LP625-44, respectively. We
expect that the rotation of these old subgiants is slow, and the
broadening represents the macro-turbulence in the atmosphere.

\begin{figure}
  \begin{center}
    \FigureFile(120mm,80mm){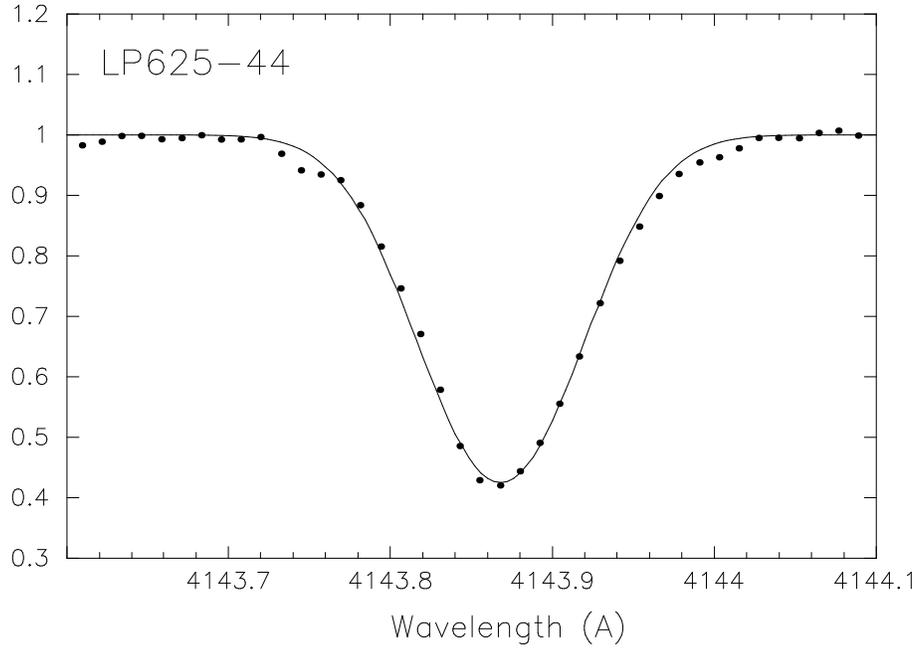}
%%% \FigureFile(width,height){filename}
  \end{center}
  \caption{Observed spectrum of LP~625-44 around the Fe {\small I}
  $\lambda 4143.8$ line (dots). The synthetic spectrum (line) fitted to
  the observed one is calculated for $v_{\rm macro}=6.5$~km~s$^{\small
  -1}$. (see text for details) 
}\label{fig:fe1}
\end{figure}

However, as found in the lower panel of Figure~\ref{fig:line} for
LP625-44, the scatter in the widths of weak lines is large, and there
seems to be an influence of blending on some weak lines. Since this
object shows remarkable overabundances of carbon, nitrogen and
neutron-capture elements, blending with absorption lines of molecules
and heavy elements is a significant problem. We tried an independent
estimation of the line broadening for a rather strong and clean Fe
{\small I} line ($\lambda$4143.87) by fitting a synthetic spectrum
(Figure~\ref{fig:fe1}). The stellar parameters used for the
calculation are given in next section. We assumed a single Gaussian
broadening here. This assumption is expected to be enough to estimate
the broadening for the following analysis, though that may not
represent the broadening process of the real atmosphere. This analysis
showed that line broadening of 7.4~km~s$^{\small -1}$ seems to be an
overestimate, but about 6.5~km~s$^{\small -1}$ is acceptable. We
adopted the later as the macro-turbulence of LP625-44.

We estimate here for completeness the effect of blending on
measurement of equivalent widths. If the discrepancy in the measured
macro-turbulence from Figure \ref{fig:line} and that from the Fe
{\small II} $\lambda$4143 line is due to blending with other
lines, that also results in the overestimate of equivalent widths for
the weak Fe {\small I} lines used in the abundance analysis, and then
the derived iron abundances. The effect is about 15\% (0.06 dex) for
some weak lines, and that does not make large impact on the abundance
determination.

\subsection{Measurement of radial velocity}\label{sec:rv}

Variation of radial velocity has been detected for LP~625-44 by
\citet{norris97a} and \citet{aoki00}. This fact indicates the
binarity, and strongly supports the mass transfer scenario as an
explanation for the overabundances of carbon and {\it s}-process
elements in this star. Since one period of the variation has not yet
been covered in the monitoring, further measurements of radial velocity
are indispensable to determine the orbital parameters for this object.

For the measurement of radial velocity, we selected clean Fe {\small
I} lines in each spectrum, and measured the line position by fitting
Gaussian profiles, and then derived the heliocentric velocity. Results
are given in Table \ref{tab:rv}, along with the standard error in each
measurement and the Julian Day (JD) of the observation. We note that
the Th-Ar comparison data were not acquired before nor after the
HD140283 observing on 18 August run, so a reliable radial velocity
could not be derived from that spectrum.

The results from the two measurements for HD~140283 agrees very
well. No variation of radial velocity is known for this object, and
our results are reasonable. The radial velocities of LP~625-44 (30.2
and 30.9~km$^{-1}$) are almost consistent with the result
(30.0~km$^{-1}$) for JD=2451569.8 \citep{aoki00}. The variation of
radial velocity for this object is expected to be small for about 200
days, and the results are also reasonable.

\section{Abundance analysis and Results}\label{sec:ana}

An abundance analysis using model atmospheres in the ATLAS grid of
\citet{kurucz93a} was made for our two metal-poor objects. While a
standard analysis based on equivalent widths was carried out for most
elements, a spectrum synthesis technique was applied to molecular
bands (CH, CN and C$_{2}$) and atomic lines which are affected by
blending and/or line splitting (hyperfine splitting and isotope
shifts). In this section, we describe the procedures and results of
the abundance analysis.
%The results of the abundances are given in table\ref{tab:res}.

\subsection{Stellar parameters}\label{sec:param}

The parameters of model atmospheres used in the abundance analysis
are effective temperature ($T_{\rm eff}$), surface gravity ($g$),
micro-turbulent velocity ($v_{\rm micro}$), and metallicity ([Fe/H] is
adopted as metallicity). Once these parameters are given, the
equivalent width is calculated for assumed elemental abundances for
each line whose lower excitation potential ($\chi$) and transition
probability ($gf$) are known. We determine the abundance from the comparison
of the calculated equivalent widths with observed ones. There are other
parameters of line broadening like macro-turbulence for the spectrum
synthesis method.

The effective temperatures have been determined by \citet{norris97a} and
\citet{ryan96} for LP625-44 (5500K) and HD140283 (5750K),
respectively. These are based on their color ($R-I$), and we adopted
these values in this analysis.

\begin{figure}
  \begin{center}
    \FigureFile(120mm,150mm){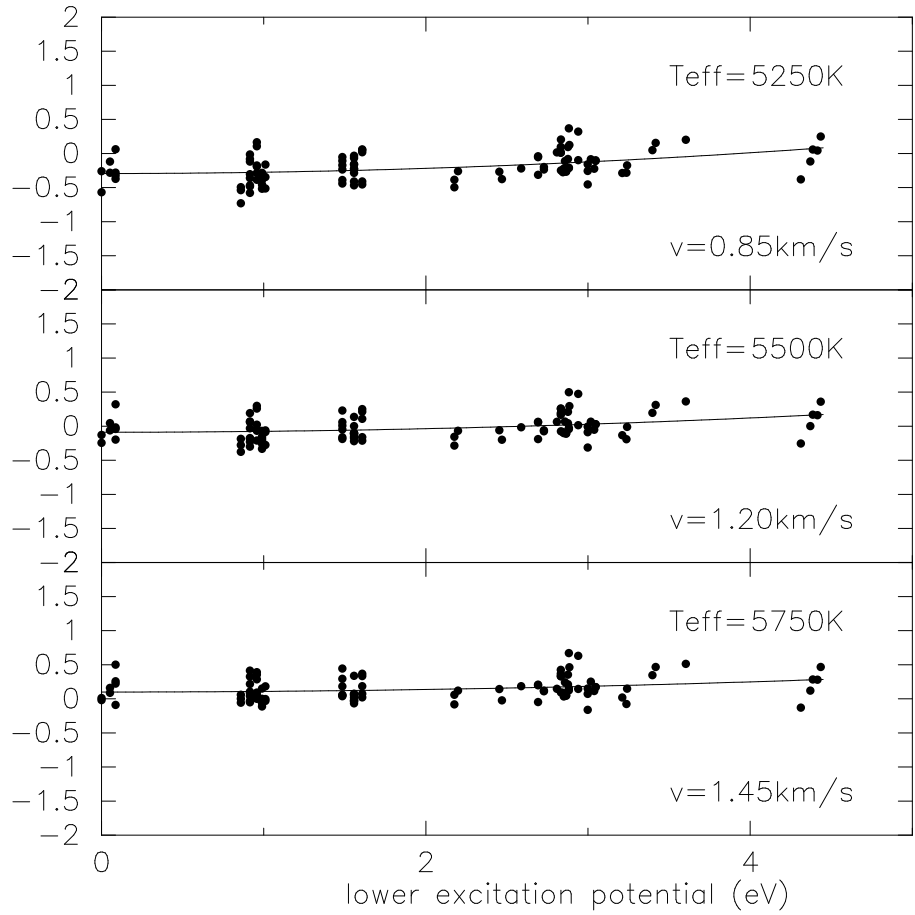}
%%% \FigureFile(width,height){filename}
  \end{center} \caption{Logarithmic abundances derived from Fe {\small
  I} lines as a function of the lower excitation potential for
  LP~625-44. The vertical axis shows the abundance difference from the
  input (assumed) one in the analysis ($\log\epsilon=4.8$). The
  effective temperature assumed and micro-turbulence ($v$) adopted in
  the analysis are given in each panel.  }\label{fig:teff}
\end{figure}

For a check of the effective temperature, we investigated the
correlation between the lower excitation potential and the resulting
abundances for Fe {\small I} lines (Figure \ref{fig:teff}). The middle
panel of this figure shows the correlation for the effective
temperature adopted in this analysis. Though the abundances resulting
from high excitation lines ($\chi \sim 4$~eV) are higher than those
from low excitation ($\chi \sim 0$~eV), the discrepancy is not
significant ($\sim$ 0.2~dex).
 
The top and bottom panels show the same ones for $T_{\rm eff}=5250$
and 5750~K, respectively. It should be noted that the abundances are
derived using $v_{\rm micro}$ re-determined for each case, instead of
the $v_{\rm micro}$ derived in the analysis for $T_{\rm
eff}=5500$K. The dependence of resulting abundances on the lower
excitation potential slightly changes, but the effect of the change of
effective temperature is not large. This is probably because the
effect of effective temperature is partly canceled out by the
re-determination of $v_{\rm micro}$, because the equivalent width is
strongly dependent on the excitation potential of the line (lines with
low excitation potential are usually strong, while high excitation
lines are usually weak). This implies that the correlation between
excitation and resulting abundance does not give a strong constraint on
the determination of effective temperature for this object. We note
there is another difficulty in the estimation of effective
temperatures from the relation between excitation and resulting
abundance due to the uncertainty in the treatment of damping
\citep{ryan98}.
 
\begin{figure}
  \begin{center} \FigureFile(120mm,150mm){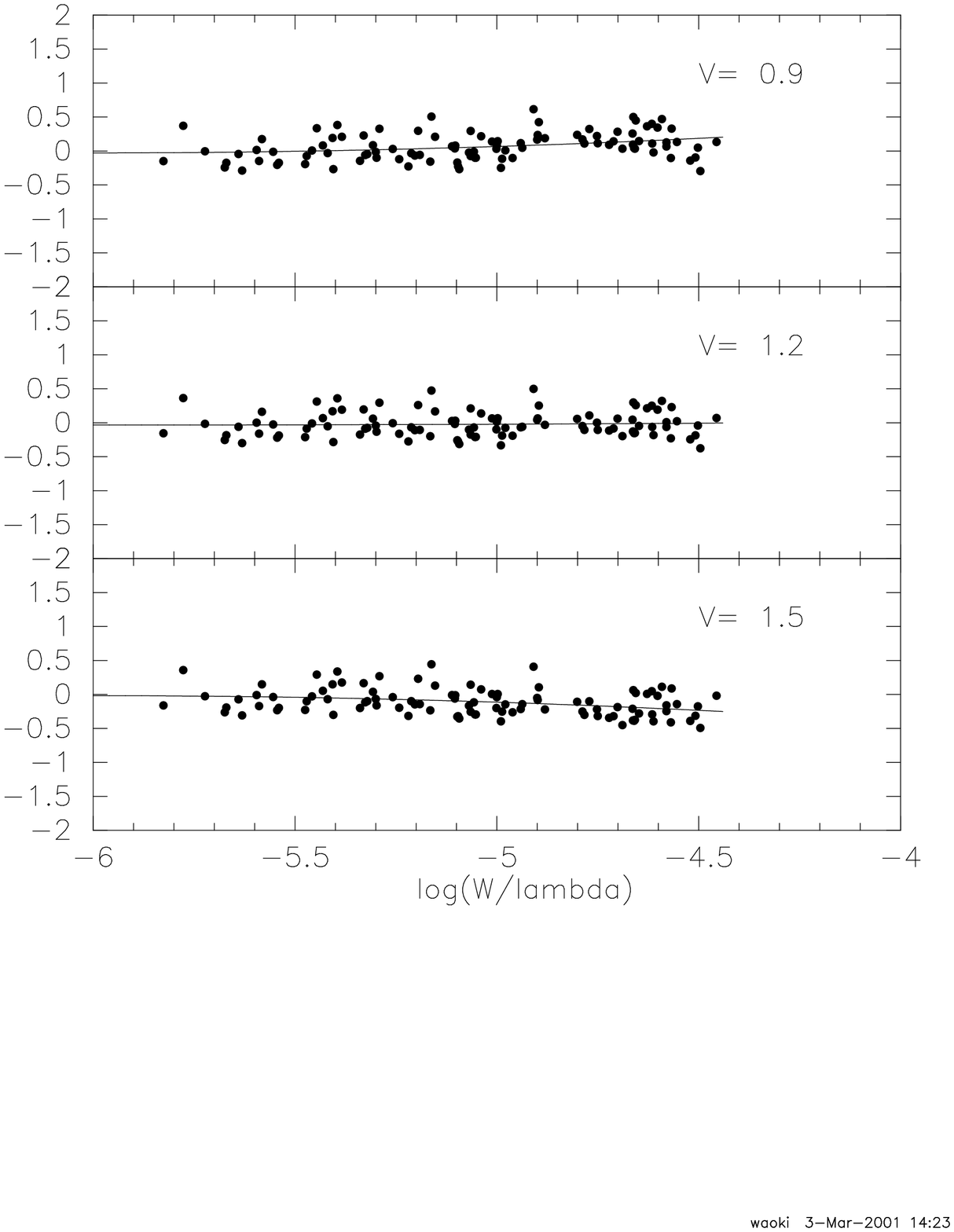} \end{center}
    \caption{Logarithmic abundances derived from Fe {\small I} lines
    as a function of equivalent width of the line for LP~625-44. The
    micro-turbulent velocity assumed (km~s$^{-1}$) is given in each
    panel.  }\label{fig:micro}
\end{figure}

The surface gravity was determined to obtain ionization balance
between Fe {\small I} and Fe {\small II} lines, and also between Ti
{\small I}/Ti {\small II}. The final derived gravities are $\log g=$
3.3 and 2.5 for HD~140283 and LP~625-44, respectively. The
micro-turbulent velocities were determined from the Fe {\small I}
lines by demanding no dependence of derived abundance on equivalent
widths. An example is shown in Figure \ref{fig:micro}. The result is
1.2~kms$^{-1}$ in both objects. These values agree rather well with
those derived by previous works \citep{ryan96, aoki00}. 

The random errors in standard analysis for Fe and Ti, which include
errors in equivalent width measurements and uncertainties of {\it
gf}-values, are about 0.1~dex or smaller. These errors result in the
uncertainties of about 0.3~dex in $\log g$ and of 0.3~km~s$^{-1}$ in
micro-turbulent velocities. We estimate the errors of the abundances
caused by the uncertainties in stellar parameters in section
\ref{sec:uncert}.

The surface gravity of HD~140283 was measured by \citet{fuhrmann98}
based on the Hipparcos parallaxes. Their result is $\log g=3.7\pm
0.1$, rather higher than the value derived from the ionization
balance by some spectroscopic studies including our present
work. The difference between the $\log g$ of \citet{fuhrmann98} and
ours (0.4~dex) is comparable to the uncertainty of $\log g$ in this
work (0.3~dex), but the discrepancy may be due to the non-LTE effect,
as suggested by \citet{fuhrmann98}. 

\subsection{Neutron-capture elements}\label{sec:heavy}

The neutron capture elements of LP~625-44 were studied by
\citet{aoki01} in detail. We re-analyzed the abundances for most
elements using the line lists produced by that work for new
spectrum. The effect of hyperfine splitting and isotope shifts were
included for the lines of Ba, La, Pr, Nd, Sm, Eu and Pb. The effect
seems very small in Ce lines, and that was neglected: see
\citet{aoki01} for details of the line lists and analysis. Here we
discuss some new lines of neutron capture elements which were not
studied by the previous work. We inspected carefully the spectrum of
HD140283, but could not find distinct absorption feature of neutron
capture elements except for Sr, Y, Zr and Ba.
 
\subsubsection{Pb I lines}\label{sec:pb}

\begin{figure}
  \begin{center}
    \FigureFile(120mm,120mm){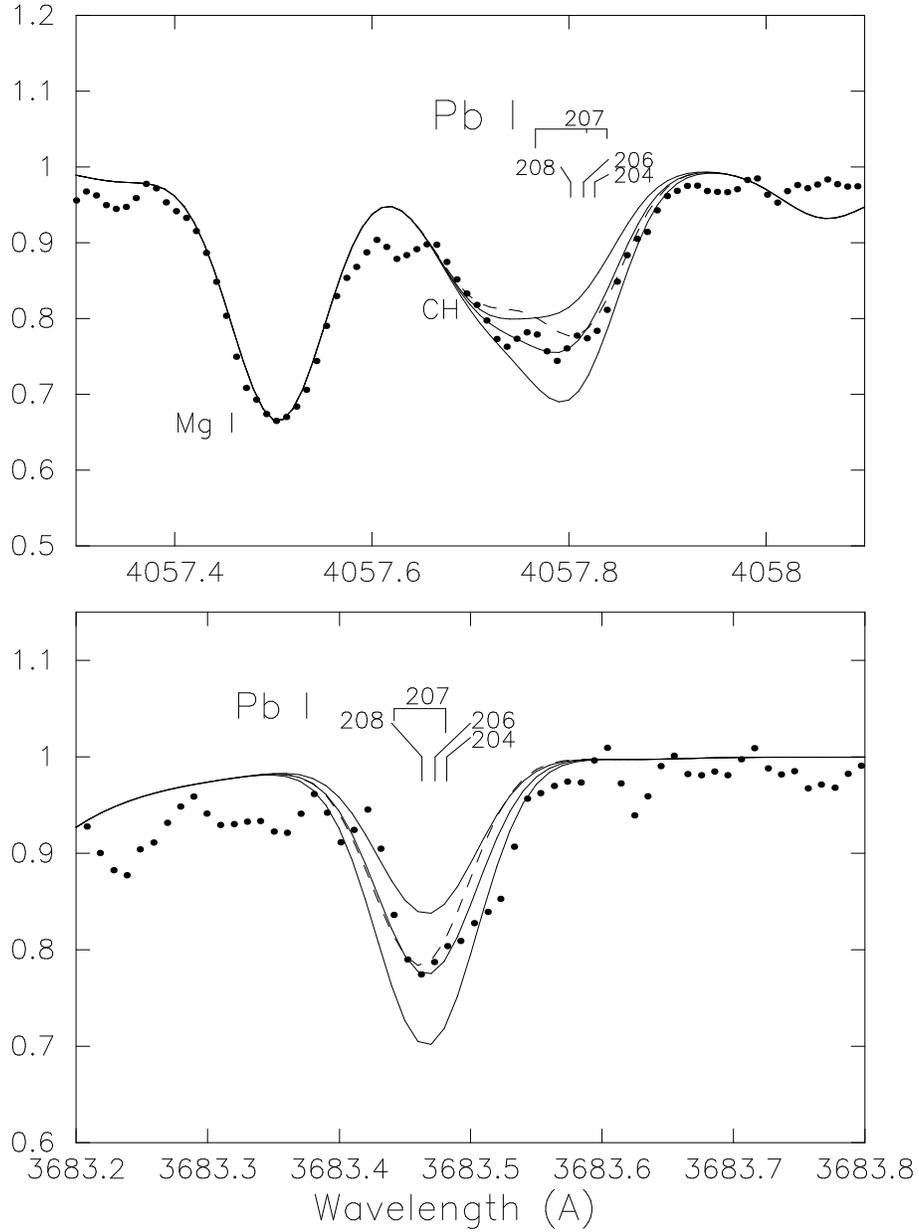}
%%% \FigureFile(width,height){filename}
  \end{center} 
\caption{ Observed spectrum of LP~625-44 for Pb {\small I} lines at
4057{\AA} (upper) and 3683{\AA} (lower). The synthetic spectra shown
by solid lines are calculated including isotope shifts and hyperfine
splitting for [Pb/Fe]=+2.6$\pm$0.2. The line positions of four Pb
isotopes are shown in each panel. The dashed lines indicate the
synthetic spectra calculated using the single-line approximation.}
\label{fig:pb}
\end{figure}

Since the Pb abundance derived by \citet{aoki00} and \citet{aoki01} is
based on the $\lambda 4057$ line alone, confirmation of the abundance
from other lines is desirable. In the solar spectrum, four Pb {\small
I} lines have been identified at 3639.5, 3683.4, 3739.9 and 4057.8
{\AA} \citep{youssef89}. Two of them ($\lambda 3683.4$ and $\lambda
4057.8$) are found in our new spectrum of
LP~625-44. Figure~\ref{fig:pb} shows the observed spectrum at the
wavelength of these two lines with the synthetic spectra. In the
calculation of the spectra, the effects of isotope ($^{204}$Pb,
$^{206}$Pb, $^{207}$Pb and $^{208}$Pb) shifts and hyperfine splitting
for $^{207}$Pb are taken into account. The line positions taken from
\citet{manning50} are shown for each component in the figure. We
assumed the solar system isotope ratio for Pb in this analysis. The
upper panel of this figure shows the same line ($\lambda 4057.8$) at
the top panel of Figure 2 in \citet{aoki01}, and the resulting
abundance of this work ($\log\epsilon({\rm Pb})=+1.9$) agrees with the
previous result. One advantage of our spectrum is its higher spectral
resolution and finer sampling. That makes it possible to investigate
the effect of hyperfine structure and isotope shifts on the line
profile which do not clearly appear in the spectrum studied by the
previous work. The dashed line in Figure \ref{fig:pb} indicates the
synthetic spectrum calculated by single-line approximation for
$\log\epsilon({\rm Pb})=+1.9$. Even if a higher Pb abundance is
assumed, the single-line approximation can never reproduce the wide
absorption feature at 4057.8{\AA}. Though the assumption of the solar
isotope ratio is adequate to derive the Pb abundance, more detailed
analysis for this line using a higher quality spectrum may provide
some information on the isotope ratios and give constraints on Pb
production.

The lower panel of Figure~\ref{fig:pb} shows the newly detected Pb
{\small I} line at 3683.5{\AA}. Though the quality of the spectrum at
this wavelength is poorer than at 4057{\AA}, the analysis is easier
because blending is not severe; this line is clearly seen even in the
solar spectrum. We note that the broad absorption in the shorter
wavelength region is the wing of the hydrogen line (H20) at
3682.8{\AA}. The abundance derived from this line ($log\epsilon({\rm
Pb})=+1.9$) agrees well with that from the $\lambda 4057.8$ line,
confirming the Pb abundance derived by the previous works. The effect
of isotope shifts and hyperfine splitting was also included in this
analysis, but the effect is less remarkable for this line, as shown in
the figure.

\subsubsection{Dy, Er and Yb}\label{sec:dyeryb}

Dysprosium (Dy, $Z=66$) and Erbium (Er, $Z=68$) have already been
detected by \citet{aoki01} in LP~625-44, but the number of lines
studied was quite limited. In our HDS spectrum, several new lines have
been discovered in the wavelength range 3600-3700~{\AA}. We analyzed
these lines using the transition probabilities taken from
\citet{kusz92} and \citet{musiol83} for Dy {\small II} and Er {\small
II}, respectively.

In our new spectrum of LP~625-44, the Ytterbium (Yb, $Z=70$) {\small
II} $\lambda 3694.2$ line has been detected. We tried to derive the
abundance by spectrum synthesis adopting the line data used in
\citet{sneden96}. Since there is no useful data on hyperfine splitting,
at least to our knowledge, the effect was not included in the
analysis. This may cause a significant overestimate of the abundance,
because this element has seven stable isotopes, and two of them
($^{171}$Yb and $^{173}$Yb) have odd neutron number. The Yb abundance
derived above by single line approximation is much higher than those
of Er ($Z=68$) and Hf ($Z=72$), but we suspect this is not real and
due to the effect of the line splitting.

\subsection{Light elements}\label{sec:light}

\subsubsection{Carbon and nitrogen}\label{sec:cn}

\begin{figure}
  \begin{center}
    \FigureFile(120mm,150mm){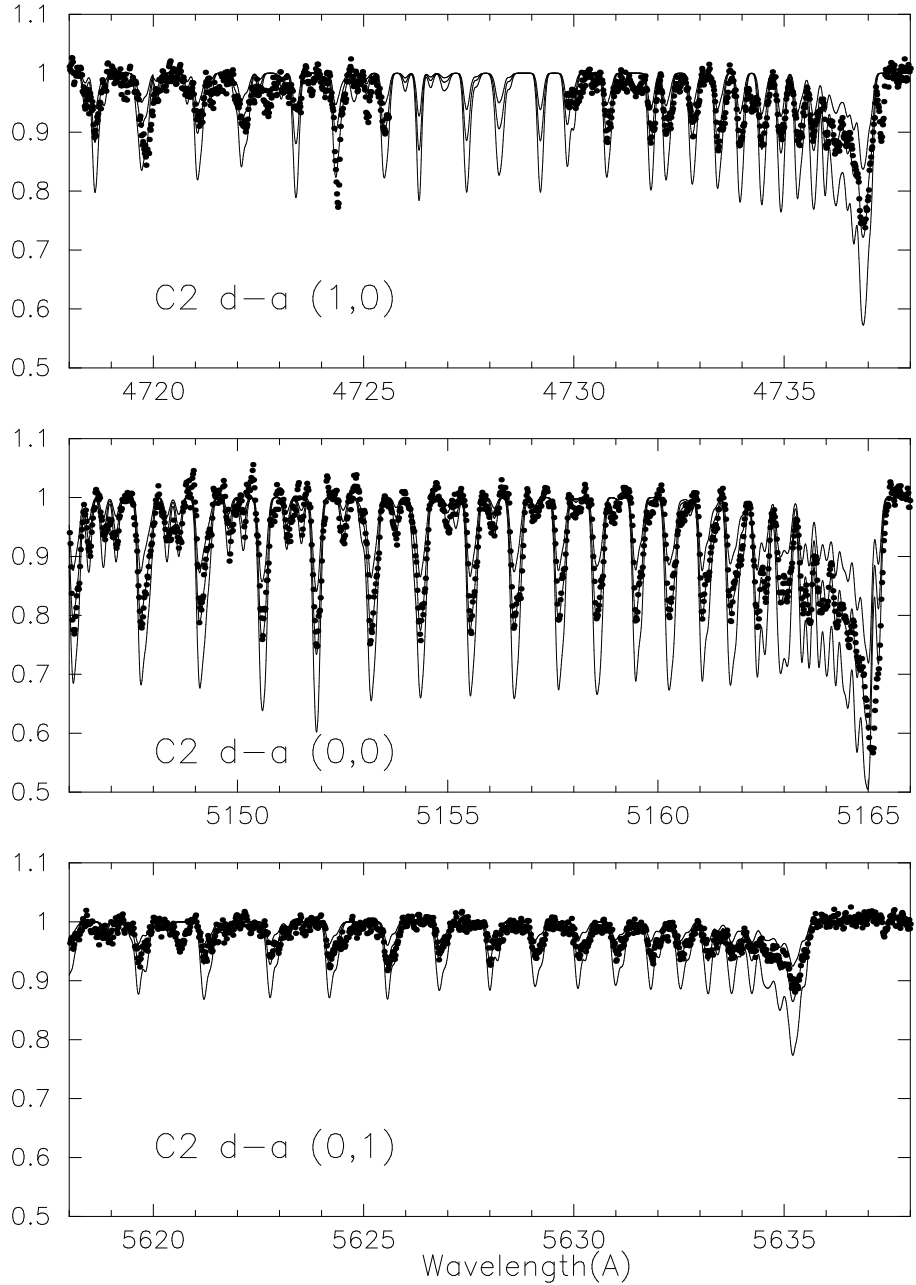}
%% \FigureFile(width,height){filename}
  \end{center} 
  \caption{
Observed spectrum of LP~625-44 for three C$_{2}$ Swan bands
(dots). The central synthetic spectra (lines) are calculated for
[C/Fe]=2.37, 2.37 and 2.47 for top, middle and bottom panels,
respectively. The other spectra are calculated for $\Delta$[C/Fe]=$\pm
0.15$ difference. The observed spectrum around 4727{\AA} is excluded
because that is affected by a bad column in the detector.
}\label{fig:c2}
\end{figure}

The carbon abundance of LP~625-44 was derived from the CH
($\lambda$4323) band by \citet{aoki01}. We first applied these
analyses to our two objects. The carbon abundance ratio derived from
the CH band is [C/Fe]=0.27 and [C/Fe]=2.17 for HD~140283 and
LP~625-44, respectively. In addition to the CH band, the C$_{2}$ Swan
bands also appear in the spectrum of LP~625-44. Figure \ref{fig:c2}
shows three bands of the Swan system of this object and the synthetic
spectra fitted to the observed one. The derived abundances are
[C/Fe]=2.37, 2.37 and 2.47 for 1-0, 0-0 and 0-1 bands of the system,
respectively. These abundances are systematically higher than that
derived from the CH band, but the agreement is acceptable giving the
uncertainty of the oscillator strengths of these molecular lines. We
adopted carbon abundance [C/Fe]=2.25 for LP~625-44.

The absorption lines of neutral carbon around 7110 {\AA} have also
been detected in LP~625-44. We performed the standard analysis for
these lines assuming LTE, and derived [C/Fe]=2.7. However, the
abundance determined from these lines may not be reliable, because the
excitation potentials of these lines are quite high ($\sim 8.6$~eV),
and the resulting abundance is quite sensitive to the model atmosphere
and adopted effective temperature. \citet{tomkin92} studied carbon
abundances for halo dwarfs based on C {\small I} lines at 9100~{\AA},
which are also high excitation (7.48~eV) lines like those at
7100~{\AA}, and found that the carbon abundances derived from the C
{\small I} lines are considerably higher ($\sim 0.4$~dex) than those
from CH molecular lines. In their study the non-LTE effect on the
abundance determination was investigated for the C {\small I} lines,
but that is not significant ($<0.1$~dex) for objects with $T_{\rm eff}
\lesssim 5500K$. Our result indicates a similar discrepancy between
the carbon abundances derived from C {\small I} lines and those from
the CH molecular band. \citet{tomkin92} also found a correlation
between carbon abundances from the C {\small I} lines and effective
temperatures, while there is no temperature dependence for carbon
abundances based on CH lines. They pointed out possible inadequacies
in the non-LTE analysis and/or model atmosphere used in the analysis
for C {\small I} lines. Considering these problems in abundance
analysis using high excitation C {\small I} lines, we adopted the
carbon abundances derived from molecular lines as mentioned above.

In the spectrum of LP~625-44, the absorption band of CN violet system
($\lambda 3883$) was also detected. Following \citet{aoki01} we
determined the nitrogen abundance ([N/Fe]=0.95) from this
band. However, it should be noted that the uncertainty in nitrogen
abundance derived from CN bands is large, as discussed in
\citet{aoki01} in detail.

\subsubsection{Oxygen}\label{sec:o}

\begin{figure}
  \begin{center}
    \FigureFile(120mm,140mm){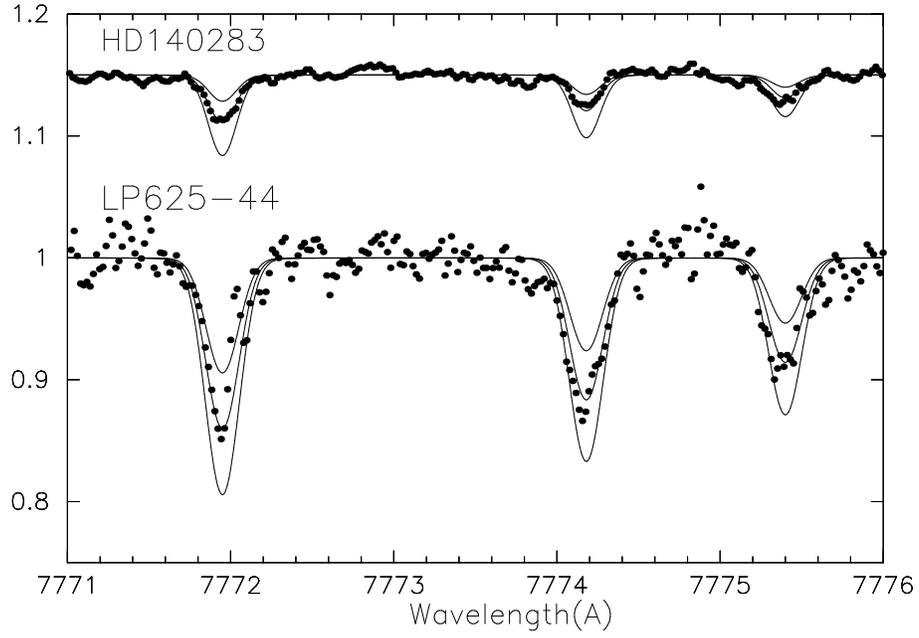}
%%% \FigureFile(width,height){filename}
  \end{center} 
  \caption{ 
Observed spectra for O {\small I} triplet (dots). The spectrum of
HD~140283 is vertically shifted by 0.15. The synthetic spectra (lines)
are calculated for [O/Fe]=0.95$\pm$0.3 and 1.85$\pm$0.3 for HD~140283
and LP~625-44, respectively.
}\label{fig:o}
\end{figure}

The oxygen triplet around 7770{\AA} has also been detected in both
objects. The observed spectra at this wavelength are shown in Figure
\ref{fig:o} (dots). The standard analysis was performed for these
lines. The derived abundances are [O/Fe]=+0.91 and +1.85 for HD~140283
and LP~625-44, respectively. We note that the absorption features of
$\lambda$7775.4 in HD~140283 and $\lambda$7772.0 in LP~625-44 seem to
be affected by some noise, like a cosmic-ray event or fringes, and
these lines are excluded in the standard analysis. In the figure,
synthetic spectra calculated for the derived oxygen abundances and
those changed by $\Delta$[O/Fe]=$\pm$0.3~dex are shown (lines).

The oxygen abundances in metal-poor dwarfs and subgiants are now in
controversy. The abundances derived from these triplet lines are
usually higher than those from the $\lambda$6300 and $\lambda$6363
forbidden lines in metal-poor dwarfs and subgiants (e.g., McWilliam
1997). For instance, the oxygen abundance of HD~140283 derived in this
analysis ([O/Fe]=+0.91) is consistent with the expected value for
[Fe/H] $\sim -2.5$ from the trend of oxygen abundances derived from
the triplet lines (e.g.,
\authorcite{boesgaard99}\yearcite{boesgaard99}). However, our oxygen
abundance is higher by about 0.5~dex than the one expected from
[O{\small I}] lines. In fact, though the [O {\small I}] lines have not
been detected in our objects, the upper limits on the oxygen
abundances estimated from the [O {\small I}] $\lambda$6300 line
(3$\sigma$ level of equivalent widths evaluated from the S/N ratios at
the wavelength) are [O/Fe]=0.76 and 1.35 for HD~140283 and LP~625-44,
respectively, lower than the oxygen abundances determined by the O
{\small I} triplet lines.

The discrepancy between the oxygen abundances derived from the O
{\small I} triplet and those from [O {\small I}] lines is beyond the
scope of this paper. The important result here is that the oxygen
abundance of LP~625-44 is obviously much higher (perhaps by nearly
1.0~dex) than that of HD~140283.

\subsubsection{Sodium}\label{sec:na}

\begin{figure}
  \begin{center}
    \FigureFile(120mm,140mm){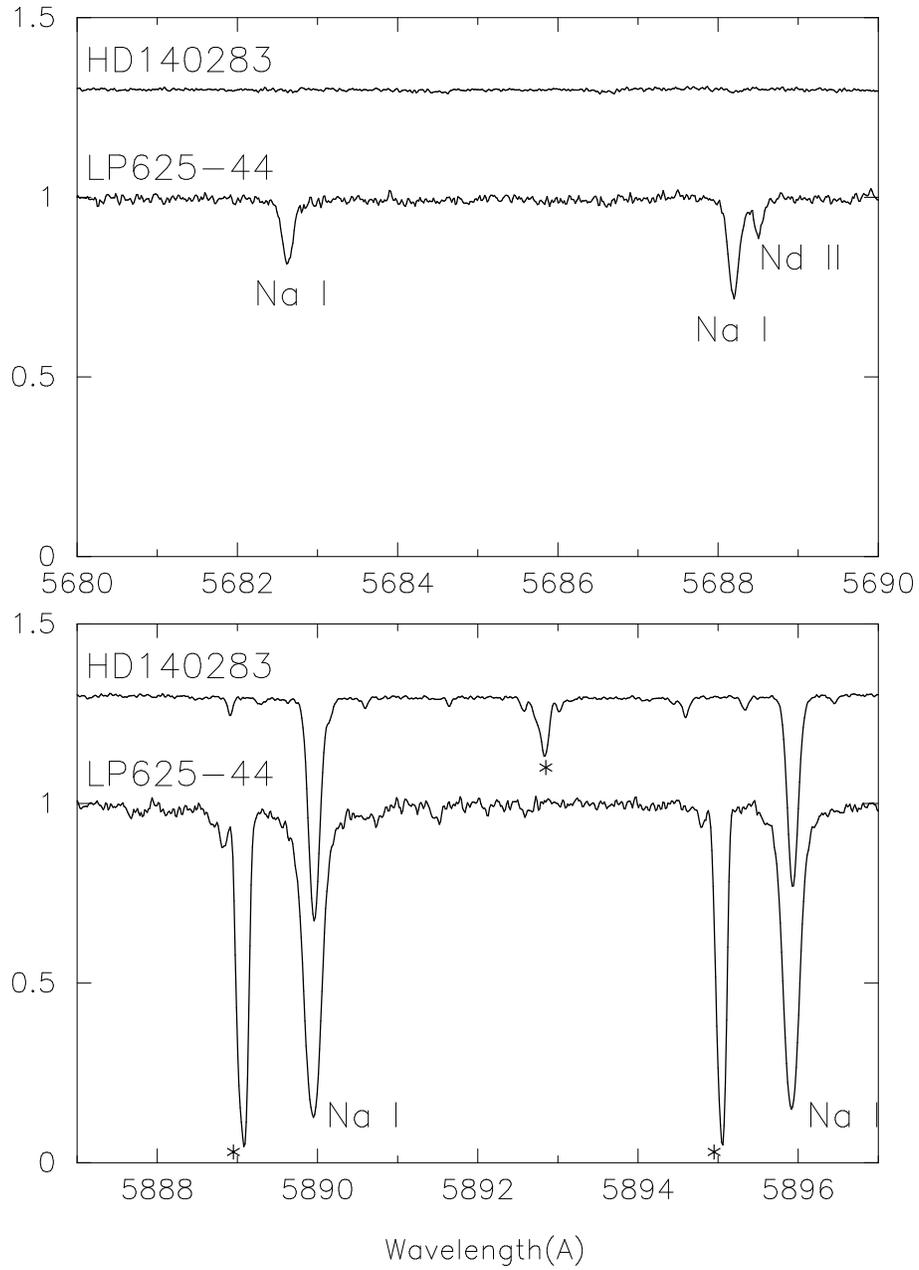}
  \end{center} 
\caption{ Observed spectra around Na {\small I} lines. The spectrum of
HD~140283 is vertically shifted by 0.3. The lines shown by asterisks
are those due to interstellar absorption. The Na {\small I} lines in
LP~625-44 are much stronger than those in HD~140283, indicating a
large overabundances of Na in LP~625-44.  }\label{fig:na}
\end{figure}

Figure \ref{fig:na} shows the observed spectra around the Na {\small
I} lines. The Na {\small I} D lines in LP~625-44 are clearly stronger
than those of HD~140283 (lower panel), and the doublet lines with
higher excitation around 5685{\AA}, which do not appear in HD~140283,
are detected in LP~625-44. This fact indicates a large excess of Na in
LP~625-44. The standard analysis for Na lines derived [Na/Fe]=+1.75
for LP~625-44 and [Na/Fe]=+0.01 for HD~140283. The Na abundance of
LP~625-44 is based on five weak lines as well as strong D lines,
implying the result is quite reliable. This remarkable overabundance
of Na in LP~625-44 will give a new constraint on the study of the
process which produces the abundance pattern of this carbon-rich
object (see section 4.2).

\subsection{Uncertainties}\label{sec:uncert}

The internal errors in the abundance determination were estimated by
the scatter (standard error) of the abundances derived from individual
lines. When the number of lines used in the abundance analysis is too
small, we assumed the random error to be 0.2~dex, which is almost the
largest scatter in the above analysis. The errors in the abundance
determination from the uncertainties of the atmospheric parameters
were evaluated by adding in quadrature the individual errors
corresponding to $\Delta T_{\rm eff}=100$K, $\Delta \log g=0.3$, and
$\Delta v=0.5$km s$^{-1}$. The uncertainties estimated for the $\Delta
T_{\rm eff}$ and $\Delta \log g$ are usually 0.05-0.10~dex. The
uncertainty due to that of micro-turbulence is strongly dependent on
the strength of the lines used in the analysis, and sometimes as large
as 0.15-0.2~dex (e.g., for Al, Ni). The uncertainties estimated are
given in Table \ref{tab:res} for the abundance ratio ([X/Fe]), except
for iron for which the uncertainty of [Fe/H] is given. We have
discussed the difficulty of the abundance determination for O and
Yb. Since the uncertainties of the abundances for these elements
cannot be estimated by the above procedure, those are not given in the
table.

\subsection{Comparison with previous works}\label{sec:comp}

The abundances of HD~140283 determined by the present study generally
agree very well with those by \citet{ryan96}. The difference of the
[Fe/H] is 0.02~dex, and those of [X/Fe] are less than 0.2~dex. An
exception is Yttrium (Y), for which we derived the abundance lower by
0.9~dex than that of \citet{ryan96}, but they commented that their
abundance of Y for this object was questionable.

The carbon abundance of HD~140283 was determined by
\citet{ryan91}. The carbon abundance determined by our analysis
([C/Fe]=0.27) is consistent with their result ([C/Fe]=0.4) within the
uncertainty. The oxygen abundance of HD~140283 was studied by
\citet{boesgaard99}. Their result derived from the O {\small I}
triplet lines adopting $T_{\rm eff}$=5694~K, which is similar to ours,
is [O/Fe]=+0.86, and the agreement is fairly good.
  
There is no systematic difference between the abundances of neutron
capture elements derived here and those by \citet{aoki01} for
LP~625-44. The discrepancies are rather large (0.3-0.4~dex) in Dy, Er
and Hf. The numbers of lines studied by \citet{aoki01} for these
elements were quite limited. We expect the present study improves the
abundance determination for these elements by increasing the number of
lines used in the analysis. The carbon and nitrogen abundances derived
here for LP~625-44 agree with those by \citet{aoki01} within the
uncertainties.

For other elements of LP~625-44, we found generally good agreement
between the abundances derived here and those by
\citet{norris97a}. Exceptions are Sc and Ni, for which
\citet{norris97a} derived by 0.8-0.9~dex higher abundances than
ours. They used only one Sc {\small II} line and two Ni {\small I}
lines in their analysis. We have not used these lines because the
lines are not covered by our spectra or seem to be affected by
blending. Instead we used other 3 Sc lines and 9 Ni lines which are
clearly detected in the bluer ($\lambda <3600${\AA}) region of our
spectra. We suspect that the previous results were affected by
blending, and believe our new result is more reliable than that by the
previous work.

\section{Discussion}\label{sec:disc}

\subsection{Overall abundance patterns}

Figure 10 shows the results of abundance analysis for HD140283 (open
boxes) and LP 625-44 (filled circles).  These two subgiants are known
to have quite similar stellar parameters to each other.  Since HD
140283 is a "normal" metal-poor subgiant which represents the typical
abundance pattern shared by the other metal-poor subgiants, it is of
great interest to compare the abundance pattern of LP 625-44 with that
of HD 140283 in order to identify several specific characteristics of
this object.

A glance at this figure tells us that the abundance ratios [X/Fe] in LP 625-44
are generally higher than those in HD 140283, except for the elements which
have 20 $\le Z \le$ 28.  Large enhancement of carbon, nitrogen, and
neutron-capture elements has already been studied in detail
(Norris, Ryan, \& Beers 1997; Aoki et al. 2001).  The present study has
confirmed several findings in the previous analyses. In addition, we found
for the first time a remarkable excess of O, Na, Mg, and Al in LP 625-44.
We will discuss characteristics of these light elements below.

%Only the exception is a very similar abundance pattern for the
%elements with 20 $\le Z \le$ 28.  

The abundance patterns of elements with 20 $\le Z \le$ 28 in LP~625-44
and HD~140283 are very similar to each other, and are typical of
metal-poor stars in the same metallicity range. Compared with the
solar system abundance ratios, Ca and Ti are moderately overabundant
by $\sim$ 0.3 dex, and Cr and Mn are rather underabundant by 0.2
$\sim$ 0.5 dex, while Co is overabundant by $\sim$ 0.2 dex.

\subsection{Oxygen and sodium overabundances}

Goriely \& Mowlavi (2000) have recently carried out a parametric study
of the {\it s}-process nucleosynthesis including the production of the light
elements.  Their model adopts one parameter of partial mixing of
protons from the hydrogen-rich envelope into the carbon-rich layers
during the third dredge-up. Protons which are supplied into the
carbon-rich layer produce $^{13}$C, which is the major neutron source
through the $^{13}$C($\alpha, n$)$^{16}$O process for subsequent
{\it s}-process nucleosynthesis in their model. They studied the efficiency
of producing the light elements and the {\it s}-process elements by adopting
various proton abundances in the partial mixing zone.

When the proton mixing is extremely strong ($X_p^{mix} > 10^{-2}$ in
their model), hydrogen burning occurs in the $^{12}$C-rich and
$^{22}$Ne-rich layer, and produces large overabundances of $^{14}$N
and $^{23}$Na as shown in Figure 1 in Goriely \& Mowlavi (2000).  The
expected process for the production of enriched $^{23}$Na in LP 625-44
is the proton capture by $^{22}$Ne, i.e. $^{22}$Ne($p,
\gamma$)$^{23}$Na.  However, if this is the case, the hydrogen burning
does not produce $^{13}$C efficiently in order to provide enough free
neutrons for the {\it s}-process.  This leads to a difficulty in
finding strongly enhanced {\it s}-process elements in this star as
shown in Figure 10.

The moderately proton-rich case ($X_p^{mix} \sim 10^{-2}$) is likely
to explain large $^{13}$C excess, which can provide enough neutrons
for the {\it s}-process nucleosynthesis, at the risk of losing
$^{23}$Na because the proton capture by $^{22}$Ne does not occur
efficiently.  Moderate enhancement in $^{16}$O and $^{24}$Mg are
expected, which is in reasonable agreement with the abundances derived
from our analysis of LP 625-44.

Having these discussions, we are now forced to speculate that the
abundance pattern of LP 625-44 may be explained in the hybrid model
which exhibits both a high proton-density zone and a moderate
proton-density zone. Physical processes which induce the proton mixing
are still unknown, although diffusion effects and stellar rotation are
possible candidates (Goriely \& Mowlavi 2000, and references
therein). Our result gives a new, strong constraint on such types of
proton mixing models of the AGB evolution.

It is worthwhile studying another scenario of the {\it s}-process
nucleosynthesis in AGB stars which has recently been proposed by
Iwamoto et al. (2001), that is a model which enables the proton mixing
into the helium burning layers at thermal pulses (see the next
subsection).

\subsection{Ratio between abundances of elements at second and third 
{\it s}-process peaks}

Detection of Pb I $\lambda$3683 in the present work has confirmed the
accuracy of the previously determined Pb abundance [Pb/Fe] = 2.6 (Aoki
et al.  2000; Aoki et al. 2001).  Pb is a unique, accessible element
among those at the third abundance peak, while there are several
elements, Sr-Y-Zr at the first abundance peak and Ba-to-Nd at the
second abundance peak, which previously been detected in several
metal-deficient stars.  We discuss in this subsection the ratios of
the abundances of the second-peak elements to the abundance of Pb,
which have shown a large difference among several recently observed
{\it s}-process-element-rich stars.

We here define an average abundance ratio of elements in the second
{\it s}-process peak to those in the third {\it s}-process peak by
[hs/Pb] = ([La/Fe]+[Ce/Fe]+[Nd/Fe])/3 $-$ [Pb/Fe], where 'hs' means
'heavy' {\it s}-process elements as used in some previous papers.
This ratio turns out to be [hs/Pb] = $-$0.36 for LP 625-44 (this
work), and $-$0.41 for LP 706-7 (Norris, Ryan \& Beers 1997; Aoki et
al. 2000).  These values are similar to each other.  Recently, van Eck
et al. (2001) have studied another three {\it s}-process-element-rich
stars HD187861 ([Fe/H] = $-$1.65), HD224959 ([Fe/H] = $-$1.7), and
HD196944 ([Fe/H] = $-$2.45) which exhibit considerably smaller
abundance ratios [hs/Pb] = $-$1.14, $-$1.13, and $-$1.23,
respectively.  We plot the [hs/Pb] values as a function of metallicity
[Fe/H] for all these five stars in Figure 11.

Although the sample of metal-poor, {\it s}-process element-rich stars
is still too small to infer some definite conclusion, one speculative
possibility is that a different nucleosynthesis condition might
produce a large scatter in the abundance ratios of {\it s}-process
elements.  The {\it s}-process in low metallicity AGB stars has ever
been studied theoretically by Gallino et al. (1998) and Goriely \&
Mowlavi (2000), based on the scenario of {\it s}-process in the
radiative layer between two thermal pulses. In these theoretical
models they introduced one adjustable parameter to treat the proton
supply to the $^{12}$C-rich layer where $^{13}$C is produced as the
major neutron source for the {\it s}-process.  Since the neutron
exposure in the {\it s}-process in AGB stars depends strongly on this
parameter, a large scatter in [hs/Pb] may result from different
conditions of the adopted parameter values although the mechanism of
proton mixing is unknown in their models.

Another possibility arises from the speculation that there are at
least two types of nucleosynthesis processes to give rise to different
abundance ratios [hs/Pb] $\approx -$0.4 (for LP 625-44 and LP 706-7)
and $-$1.2 (for HD187861, HD224959, and HD196944).  Quite recently,
\citet{iwamoto02} have found theoretically that the proton mixing
into helium burning layer occurs at thermal pulses only in very
metal-deficient ([Fe/H] $\le -$2.5) AGB stars. This kind of mixing is
prevented by the large entropy barrier between the helium-rich zone
and the hydrogen-rich envelope in the stars with higher metallicity
\citep{fujimoto00}. They have also shown that the {\it s}-process
proceeds under a condition of large neutron exposure in the convective
helium burning shell so that the resultant {\it s}-process abundance
pattern fits very well to that observed in two metal-deficient stars
LP 625-44 and LP 706-7 but is different from those in the three stars
studied by \citet{vaneck01}.  This theoretical prediction should
be tested by further observation of the metallicity dependence of
[hs/Pb] in the {\it s}-process elemet-rich stars.

Further studies of the {\it s}-process nucleosynthesis in metal-deficient AGB
stars are desired to correctly interpret the difference in [hs/Pb] found in
the present study.  It is also important to study the origin of Pb in the
solar system, which is one of the long-standing problems in the chemical
evolution of the Galaxy, by clarifying the difference from the {\it s}-process
nucleosynthesis in the metal-deficient AGB stars.  Note, however, that any
theoretical models are required to explain not only the abundance ratios of
neutron-capture elements but also several characteristics of the light
elements which were discussed in the previous subsection.

\section{Concluding remarks}\label{sec:conc}

The high resolution spectrum with wide wavelength coverage obtained
with Subaru/HDS for the carbon-rich, very metal-poor star LP~625-44
provided new constraints on the modelings for the {\it s}-process at
low metallicity and AGB evolution.  We found significant
overabundances in some light elements, in addition to carbon and
nitrogen whose enhancement has already been known. The Na enhancement
in LP~625-44 is significant, suggesting that hydrogen burning in the
$^{22}$Ne-rich layer in an AGB star produces the abundance pattern of
this object. The oxygen abundance was derived from O {\small I}
triplet around 7770{\AA}, and the excess of oxygen in LP~625-44 is
also remarkable ($\sim$ factor 10) compared with that of HD~140283,
though there are uncertainties in the oxygen abundances which are well
known in metal-poor dwarfs and subgiants. The newly detected Pb
{\small I} $\lambda$3683{\AA} line confirmed the Pb abundance derived
by the previous work from the Pb {\small I} $\lambda$4057{\AA}
line. The abundance ratio of the elements at the second peak to the
third peak of the {\it s}-process ([hs/Pb] defined in the present
work) shows a large spread (or two distinct classes) in the {\it
s}-process element-rich stars. Further study for a much larger sample
covering a wide metallicity range is crucial to investigate the
characteristics of the {\it s}-process in the early Galaxy.

The next step of the study on the {\it s}-process element-rich stars
like LP~625-44 is the measurements of the neutron densities and
temperatures during the {\it s}-process. These parameters will
distinguish the processes occurring during the thermal pulse (short
time-scale processes with high neutron density and high temperature)
and those occurring during the inter-pulse phase (long time-scale
processes with low neutron density and low temperature). The
estimation of neutron density and temperature for the processes which
have contributed to the abundances of the observing object may be
possible by the measurements of isotope ratios for some elements near
the branching points (e.g., $^{151}$Sm). The high resolving power of
HDS will provide an opportunity for isotope studies of these objects.

\bigskip

We are grateful to N. Iwamoto for fruiteful discussions on this topic.
Thanks are due to S.G. Ryan for reading the manuscript with helpful
comments.

\clearpage
%%%%%%%%%%%%%%%%%%%%%%%%%%%%%%%%%%%%%%%

%\begin{longtable}{ll}
%  \caption{Sample of long tabular}\label{tab:LTsample}
%  \hline\hline
%  name & value \\
%\endfirsthead
%  \hline\hline
%  name & value \\
%\endhead
%  \hline
%\endfoot
%  \hline
%\endlastfoot
%  aaaaa & bbbbb \\
%  ...... & ..... \\
%  yyyyy & zzzzz \\
%\end{longtable}

%\appendix
%\section{Method of .....}

%\section{Approximation of ...}

%\section*{Complete data}

%%%
% See the manual for the detail.
%

%
%
%\begin{figure}
%  \begin{center}
%    \FigureFile(120mm,100mm){yb.eps}
%%%% \FigureFile(width,height){filename}
%  \end{center} 
%  \caption{
%Observed spectrum of LP~625-44 around Yb {\small II} $\lambda 3694$
%line (dots). The synthetic spectra calculated for [Yb/Fe]= 3.1, 3.3
%and 3.5 are also shown (lines). The analysis by single line
%approximation for this line may cause significant
%overestimation of the Yb abundance (see text).
%  }\label{fig:yb2}
%\end{figure}
%%
\begin{figure}
  \begin{center}
    \FigureFile(120mm,120mm){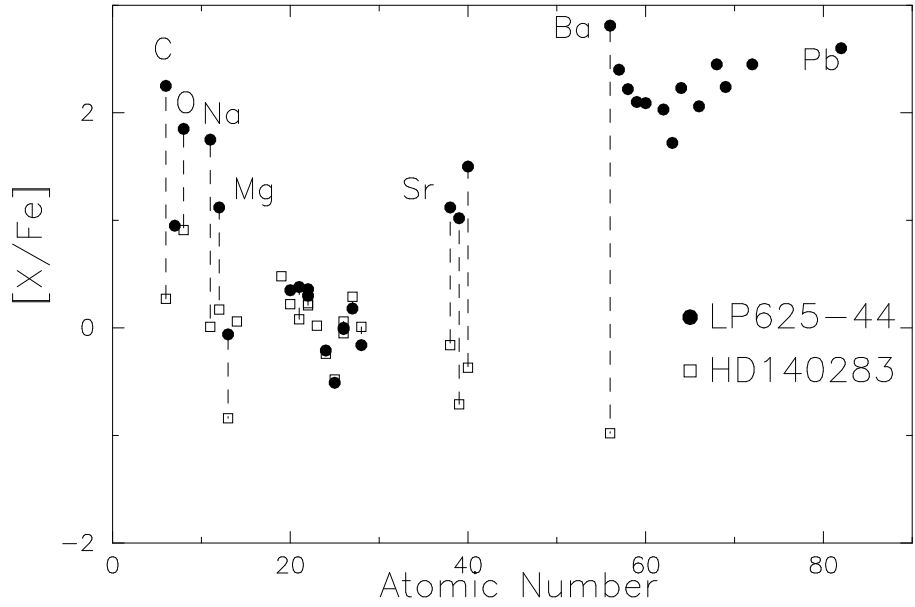}
%%% \FigureFile(width,height){filename}
  \end{center} 
  \caption{ 
Abundance ratio ([X/Fe]) as a function of atomic number.
}\label{fig:abundance}
\end{figure}
\begin{figure}
  \begin{center}
    \FigureFile(120mm,120mm){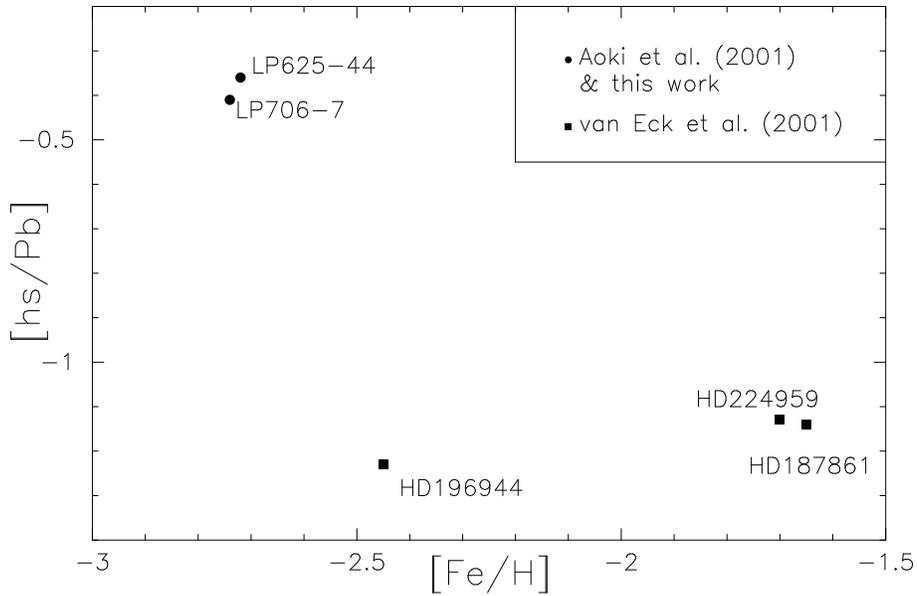}
%%% \FigureFile(width,height){filename}
  \end{center} 
\caption{Ratio between abundances of elements at second and third {\it
s}-process peaks ([hs/Pb]), as a function of [Fe/H] for {\it
s}-process element-rich stars. The abundance ratio of the second {\it
s}-process peak is defined as [hs/Fe]=([La/Fe]+[Ce/Fe]+[Nd/Fe])/3
(see text for details)}
\label{fig:hspb}
\end{figure}

%%%
\clearpage
\begin{longtable}{llccc}
\caption{Observations}\label{tab:obs}
\hline
\hline
Object & Obs. date  & Wavelength({\AA}) & exposure (sec) & S/N (Wavelength)$^{\rm a}$ \\
\endhead
\hline
LP~625-44 & 3 July, 2000 & 5100-7800 & 3600 & 130 (6000{\AA})\\
          & 21 Aug, 2000 & 3400-5100 & 3600 & 70 (4000{\AA})\\
HD~140283 & 18 Aug, 2000 & 3100-4500 & 1800 & 330 (4000{\AA})\\
          & 5 July, 2000 & 3400-5100 & 1800 & 560 (4800{\AA})\\
          & 3 July, 2000 & 5100-7800 & 960  & 680 (6000{\AA})\\
\noalign{\smallskip}
\hline 
\end{longtable}
%~\\

$^{a}$ S/N ratio per 0.9~km~s$^{-1}$ pixel.  \\

%~ \\
%
%\vspace{3cm} 
%~ \\
\newpage
%### RV
\begin{longtable}{lcccc}
\caption{Radial Velocity Measurement}\label{tab:rv}
\hline
\hline
Object & JD  & wavelength & Radial velocity & number  \\
       &     & {\AA}      & (km~s$^{-1}$) & of lines \\
\endhead
\hline
LP~625-44 & 2451728.9 & 5100-6500 & 30.92$\pm$0.26 & 26 \\
          & 2451777.8 & 3700-5100 & 30.20$\pm$0.30 & 46 \\
HD~140283 & 2451728.8 & 5100-6500 & $-$170.58$\pm$0.15 & 23 \\
          & 2451730.9 & 4500-5100 & $-$170.64$\pm$0.34 & 21 \\
\noalign{\smallskip}
\hline
\end{longtable}
\clearpage
%### EW
\begin{longtable}{rrrrrrl}
\caption[]{Results}\label{tab:ew}
  \hline\hline
Element & Wavelength ({\AA}) & L.E.P. (eV) & $\log gf$ & HD140283 & LP625-44 \\
  \hline
\endhead
\hline
      C I  &  7111.475 &   8.640 &  -1.130 &    ... &   21.0 & (1) \\
      C I  &  7113.180 &   8.650 &  -0.810 &    ... &   36.6 & (1) \\
      C I  &  7115.186 &   8.640 &  -0.710 &    ... &   32.5 & (1) \\
      C I  &  7116.990 &   8.650 &  -0.940 &    ... &   36.6 & (1) \\
      C I  &  7119.671 &   8.640 &  -1.070 &    ... &   27.4 & (1) \\
     Na I  &  4978.541 &   2.102 &  -1.220 &    ... &   12.4 & (2) \\
     Na I  &  4982.813 &   2.104 &  -0.960 &    ... &   17.2 & (2) \\
     Na I  &  5682.633 &   2.102 &  -0.700 &    ... &   33.9 & (2) \\
\hline
\end{longtable}
\clearpage
\begin{longtable}{rrrrrrl}
\caption[3]{(Continued)}\label{tab:ew}
  \hline\hline
Element & Wavelength ({\AA}) & L.E.P. (eV) & $\log gf$ & HD140283 & LP625-44 \\
  \hline
\endhead
\hline
     Fe I  &  3841.048 &   1.608 &  -0.065 &   88.5x &   93.0 & (16) \\
     Fe I  &  3843.257 &   3.047 &  -0.240 &   30.9 &   29.8 & (14) \\
     Fe I  &  3845.169 &   2.424 &  -1.390 &   16.2 &    ... & (14) \\
     Fe I  &  3846.410 &   3.573 &  -0.470 &   12.0 &    ... & (14) \\
     Fe I  &  3850.818 &   0.990 &  -1.745 &   60.0 &    ... & (16) \\
     Fe I  &  3852.573 &   2.176 &  -1.180 &   26.0 &    ... & (14) \\
     Fe I  &  3854.368 &   3.211 &  -1.100 &    5.9 &    ... & (14) \\
     Fe I  &  3856.372 &   0.052 &  -1.286 &  103.2x &   93.6 & (15) \\
\end{longtable}
\clearpage
\begin{longtable}{rrrrrrl}
\caption[3]{(Continued)}%\label{tab:ew}
  \hline\hline
Element & Wavelength ({\AA}) & L.E.P. (eV) & $\log gf$ & HD140283 & LP625-44 \\
  \hline
\endhead
\hline
     Ni I  &  5035.374 &   3.635 &   0.290 &    6.1 &    ... & (15) \\
     Ni I  &  5080.523 &   3.655 &   0.130 &    4.9 &    ... & (15) \\
     Ni I  &  5137.075 &   1.676 &  -1.990 &    8.1 &    ... & (15) \\
     Ni I  &  5476.906 &   1.826 &  -0.890 &   29.8 &    ... & (15) \\
     Ni I  &  6767.778 &   1.826 &  -2.170 &    2.1 &    ... & (15)  \\
     Sr II &  4077.714 &   0.000 &   0.150 &   76.4 &  139.5 & (21) \\
     Sr II &  4161.796 &   2.940 &  -0.500 &    ... &    8.3 & (21) \\
     Sr II &  4215.524 &   0.000 &  -0.180 &   66.0 &  115.3 & (21) \\
\hline
\end{longtable}
\clearpage
$^{\rm a}$This table is available only in electronic version. 

$^{\rm b}$ Most line data have been compiled in \citet{hirata94}.  \\

Rerefences.---
(1)\citet{tomkin95};  %Tomkin et al. 1995
(2)\citet{WM80};      %Wiese & Martin 1980
(3)\citet{morton91};   %Morton 1991
(4)\citet{joensson84}; %Joensson,G.,S.Kroll,A.Persson & S.Svanberg           *Phys.Rev. A30, 2429-32
(5)\citet{gigas88};    %Gigas,D. 1988 Astron.Astrophys. 192, 264-74   
(6)\citet{smith87};    %Smith,P.L.,M.C.Huber,G.P.Tozzi,H.E.Griesinger,B.L.Cardon & G.G.Lombardi *Astrophys.J. 322, 573-83
(7)\citet{smith81};    % Smith,G. & D.St. J. Raggett  J.Phys. B14, 4015-24 
(8)\citet{lawler89};   % Lawler,J.E. & J.T.Dakin   J.Opt.Soc.Amer. B6, 1457-66 
(9)\citet{WF75};    % Wiese,W.L. & J.R.Fuhr J.Phys.Chem.Ref.Data 4, 263-349
(10)\citet{MFW88};    %Martin, Fuhr & Wiese 1988
(11)\citet{bizzarri93}; %Bizzarri,A.,M.C.E.Huber,A.Noels,N.Grevesse,S.D.Bergeson,P.Tsekeris & J.E.Lawler *Astron.Astrophys. 273, 707
(12)\citet{savanov90}; %Savanov,I.S.,J.Huovelin & I.Tuominen Astron.Astrophys.Suppl. 86, 531
(13)\citet{biemont89a}; % Biemont,E.,N.Grevesse,L.M.Faires,G.Marsden,J.E.Lawlwer & W.Whaling *Astron.Astrophys. 209, 391
(14)\citet{obrian91};        %O'Brian et al. 1991
(15)\citet{fuhr88};    %Fuhr, Martin & Wiese 1988 
(16)\citet{bard91};   %Bard, Kock & Kock 1991
(17)\citet{pauls90}  %Pauls,U.,N.Grevesse & M.C.E.Huber *Astron.Astrophys. 231, 536
(18)\citet{HK90};    %Heise,C. & M.Kock  *Astron.Astrophys. 230, 244-7 
(19)\citet{CC83};      %Cowley & Corliss 1983
(20)\citet{BBPL89};  %Blackwell et al. 1989
(21)\citet{GBHL81};  %Grevesse,N.,E.Biemont,P.Hannaford & R.M.Lowe *In 23rd Liege Astrophys.Colloq. pp211
(22)\citet{hannaford82}; % Hannaford,P.,R.M.Lowe,N.Grevesse,E.Biemont & W.Whaling *Astrophys.J. 261, 736
(23)\citet{CB62};      %Corliss & Botzman 1962
(24)\citet{kastberg93}; %Kastberg,A.,P.Villemoes,A.Arnesen,F.Heijkenskjoeld,P.Jungner & S.Linnaeus *J.Opt.Soc.Amer. B10, 1330
(25)\citet{BBC96};   %Bord et al. 1996
(26)\citet{sneden96};  %Sneden et al. 1996
(27)\citet{G91};        %Goly et al. 1991
(28)\citet{W85};        %Ward et al. 1985
(29)\citet{biemont89b};      %Biemont 1989
(30)\citet{bergstrom88};  %Bertstrom 1988
(31)\citet{kusz92};   %Kusz 1992 
(32)\citet{musiol83};  %Musiol et al. 1983
(33)\citet{youssef89};  %

\clearpage
%### Result
\begin{longtable}{cccccccccc}
\caption[4]{Abundance Results}\label{tab:res}
  \hline\hline
 & \multicolumn{4}{c}{HD140283} & & \multicolumn{4}{c}{LP625-44} \\
\cline{2-5} \cline{7-10}
Element \hspace{2cm} & [X/Fe] & $\log\epsilon_{\rm el}$ & n & $\sigma$ & & [X/Fe] & $\log\epsilon_{\rm el}$ & n & $\sigma$ \\
\endhead
\hline
C (CH, C$_{2}$)\dotfill & $+$0.27   & 6.3  &  & 0.20 & &  $+$2.25  & 8.1 &   & 0.23 \\ 
N (CN)         \dotfill &       &   &  &      & &         $+$0.95  & 6.2 &   & 0.35 \\ 
O            \dotfill   & $+$0.91: & 7.26: & 2  &    & &  $+$1.85: & 8.00: & 2 &  \\
Na I          \dotfill  & $+$0.01 & 3.82 & 2  & 0.24 & &  $+$1.75  & 5.35 & 7 & 0.14 \\
Mg I          \dotfill  & $+$0.17 & 5.23 & 8  & 0.21 & &  $+$1.12  & 5.98 & 4 & 0.24 \\
Al I          \dotfill  & $-$0.84 & 3.11 & 2  & 0.30 & &  $-$0.06  & 3.71 & 1 & 0.25 \\
Si I          \dotfill  & $+$0.06 & 5.10 & 1  & 0.27 & &           &      &     &     \\
K I           \dotfill  & $+$0.48 & 3.09 & 2  & 0.20 & &           &      &     &     \\
Ca I          \dotfill  & $+$0.22 & 4.05 & 25 & 0.14 & &  $+$0.35  & 3.98 & 5 & 0.21 \\
Sc II         \dotfill  & $+$0.08 & 0.66 & 16 & 0.21 & &  $+$0.38  & 0.76 & 3 & 0.34 \\
Ti I          \dotfill  & $+$0.21 & 2.63 & 33 & 0.22 & &  $+$0.30  & 2.52 & 7 & 0.19 \\
Ti II         \dotfill  & $+$0.23 & 2.65 & 96 & 0.21 & &  $+$0.36  & 2.58 & 15 & 0.25 \\
V II          \dotfill  & $+$0.02 & 1.52 & 13 & 0.23 & &           &      &     &     \\
Cr I          \dotfill  & $-$0.24 & 2.93 & 18 & 0.23 & &  $-$0.21  & 2.76 & 3 & 0.26 \\
Mn I          \dotfill  & $-$0.48 & 3.47 & 10 & 0.21 & &  $-$0.51  & 2.30 & 2 & 0.27  \\
Fe I ([Fe/H]) \dotfill  & $-$2.58 & 4.92 & 494 & 0.16 & & $-$2.72  & 4.78 & 102 & 0.20 \\
Fe II ([Fe/H]) \dotfill & $-$2.47 & 5.03 & 38 & 0.21 & &  $-$2.73  & 4.77 &  7 & 0.15 \\
Co I          \dotfill  & $+$0.29 & 2.68 & 57 & 0.28 & &  $+$0.18  & 2.37 & 17 & 0.19 \\
Ni I          \dotfill  & $+$0.01 & 3.74 & 53 & 0.27 & &  $-$0.16  & 3.37 & 9 & 0.27 \\
Sr II \dotfill          & $-$0.16 & 0.24 & 2  & 0.33 & &  $+$1.12  & 1.32 & 3  & 0.24 \\
Y II  \dotfill          & $-$0.71 & $-$1.00 & 9 & 0.22 && $+$1.02  & 0.53 & 13 & 0.22 \\
Zr II \dotfill          & $-$0.37 & 0.10 & 12 & 0.22 & &  $+$1.50  & 1.39 & 15 & 0.21\\
Ba II \dotfill          & $-$0.98 & $-$1.28 &1& 0.24 & &  $+$2.81  & 2.31 & 4  & 0.24  \\
La II \dotfill          &         &      &    &    & &    $+$2.40  & 0.90  & 13  & 0.26 \\
Ce II \dotfill          &         &      &    &    & &    $+$2.22  & 1.14  & 33  & 0.23 \\
Pr II \dotfill          &         &      &    &    & &    $+$2.10  & 0.18  & 6  & 0.26 \\
Nd II \dotfill          &         &      &    &    & &    $+$2.09  & 0.86  & 10  & 0.17 \\
Sm II \dotfill          &         &      &    &    & &    $+$2.03  & 0.29  & 10 & 0.22 \\
Eu II \dotfill          &         &      &    &    & &    $+$1.72  & $-0.48$ & 2  & 0.24 \\
Gd II \dotfill          &         &      &    &    & &    $+$2.23  & 0.60  & 7  & 0.17 \\
Dy II \dotfill          &         &      &    &    & &    $+$2.06  & 0.51  & 8  & 0.25 \\
Er II \dotfill          &         &      &    &    & &    $+$2.45  & 0.70  & 6  & 0.27 \\
Tm II \dotfill          &         &      &    &    & &    $+$2.24  & $-0.33$  & 3 & 0.27   \\
Yb II \dotfill          &         &      &    &    & &    $+$3.36: & 1.60: & 1  &   \\
Hf II \dotfill          &         &      &    &    & &    $+$2.45  & 0.48  & 2  & 0.27 \\
Pb I  \dotfill          &         &      &    &    & &    $+$2.6   & 1.9   & 2  & 0.22 \\
\hline
\end{longtable}
\end{document}